\documentclass[amssymb, superscriptaddress, twocolumn, aps, prl,longbibliography,nobibnotes]{revtex4-2}
\usepackage{graphicx}
\usepackage{amsmath}
\usepackage{upgreek}
\usepackage{color}
\usepackage{xcolor}
\definecolor{navy}{rgb}{0,0,0.64}
\usepackage{setspace}
\usepackage{hyperref}
\hypersetup{colorlinks,breaklinks,
            urlcolor=[rgb]{0,0,0.64},
            linkcolor=[rgb]{0,0,0},
            citecolor=[rgb]{0,0,0.64}}
\usepackage{nccmath}
\usepackage{amssymb}
\usepackage{bm}
\usepackage{bbding}
\setcitestyle{super}
\usepackage{multirow}
\usepackage{makecell}
\usepackage[T2A]{fontenc}
\usepackage[utf8]{inputenc}

\newcommand{\KLLP}{Key Laboratory for Laser Plasmas (Ministry of Education), School of Physics and Astronomy, Shanghai Jiao Tong University, Shanghai 200240, China.}
\newcommand{\CICIFSA}{Collaborative Innovation Center of IFSA (CICIFSA), Shanghai Jiao Tong University, Shanghai 200240, China.}
\newcommand{\Cal}{Department of Chemistry, University of California at Berkeley, Berkeley, California 94720, USA.}
\newcommand{\LBNL}{Materials Sciences Division, Lawrence Berkeley National Laboratory, Berkeley, California 94720, USA.}
\newcommand{\UCLA}{Department of Physics and Astronomy, University of California at Los Angeles, Los Angeles, California 90095, USA.}
\newcommand{\BNL}{Condensed Matter Physics and Materials Science Division, Brookhaven National Laboratory, Upton, New York, 11973, USA.}
\newcommand{\SHTech}{School of Physical Science and Technology, ShanghaiTech University, Shanghai 201210, China.}
\newcommand{\SHTechTP}{ShanghaiTech Laboratory for Topological Physics, Shanghai 201210, China.}
\newcommand{\TDLI}{Tsung-Dao Lee Institute, Shanghai Jiao Tong University, Shanghai 200240, China.}
\newcommand{\ZIAS}{Zhangjiang Institute for Advanced Study, Shanghai Jiao Tong University, Shanghai 200240, China.}
\newcommand{\Amsterdam}{Institute for Theoretical Physics Amsterdam, University of Amsterdam, Science Park 904, 1098 XH Amsterdam, The Netherlands.}

\newcommand{\papertitle}{Ultrafast formation of topological defects in a 2D charge density wave}

\begin{document}

\thispagestyle{empty}

\onecolumngrid
\begin{center}
\textbf{\Large \papertitle}
\vspace{0.2cm}
\end{center}

\begin{center}
    
Yun~Cheng,$^{1,\,2,\,{\color{navy}\ast}}$ Alfred~Zong,$^{3,\,4,\,{\color{navy}\ast},\,{\color{navy}\text{\Envelope}}}$ Lijun~Wu,$^{5}$ Qingping~Meng,$^{5}$ Wei~Xia,$^{6,\,7}$ 
Fengfeng~Qi,$^{1,\,2}$ Pengfei~Zhu,$^{1,\,2}$ Xiao~Zou,$^{1,\,2}$ Tao~Jiang,$^{1,\,2}$ Yanfeng~Guo,$^{6}$ 
Jasper~van~Wezel,$^{8}$ Anshul~Kogar,$^{9}$ Michael~W.~Zuerch,$^{3,\,4,\,{\color{navy}\text{\Envelope}}}$ 
Jie~Zhang,$^{1,\,2,\,10,\,{\color{navy}\text{\Envelope}}}$
Yimei~Zhu,$^{5,\,{\color{navy}\text{\Envelope}}}$ Dao~Xiang$^{1,\,2,\,10,\,11,\,{\color{navy}\text{\Envelope}}}$ 

\vspace{0.3cm}

(Dated: \today)
\end{center}

\vspace{0.15cm}

\begin{small}
\begin{singlespace}
{\it
\noindent$^1$\KLLP
\vspace{0.1cm}

\noindent$^2$\CICIFSA
\vspace{0.1cm}

\noindent$^3$\Cal
\vspace{0.1cm}

\noindent$^4$\LBNL
\vspace{0.1cm}

\noindent$^5$\BNL
\vspace{0.1cm}

\noindent$^6$\SHTech
\vspace{0.1cm}

\noindent$^7$\SHTechTP
\vspace{0.1cm}

\noindent$^8$\Amsterdam
\vspace{0.1cm}

\noindent$^9$\UCLA
\vspace{0.1cm}

\noindent$^{10}$\TDLI
\vspace{0.1cm}

\noindent$^{11}$\ZIAS
\vspace{0.1cm}
}

\noindent$^{\color{navy}\ast\,}$These authors contributed equally to this work: Yun~Cheng and Alfred~Zong.
\vspace{0.1cm}

\noindent$^{\color{navy}\text{\Envelope}\,}$Correspondence: A.Z. (\href{mailto:alfredz@berkeley.edu}{alfredz@berkeley.edu}), M.W.Z. (\href{mailto:mwz@berkeley.edu}{mwz@berkeley.edu}), J.Z. (\href{mailto:jzhang1@sjtu.edu.cn}{jzhang1@sjtu.edu.cn}), Y.Z. (\href{mailto:zhu@bnl.gov}{zhu@bnl.gov}), and D.X. (\href{mailto:dxiang@sjtu.edu.cn}{dxiang@sjtu.edu.cn}).
\end{singlespace}
\end{small}
\vspace{0.4cm}

\twocolumngrid

\noindent\textbf{Topological defects play a key role in nonequilibrium phase transitions, ranging from birth of the early universe \cite{Zurek1985} to quantum critical behavior of ultracold atoms \cite{Keesling2019}. In solids, transient defects are known to generate a variety of hidden orders not accessible in equilibrium \cite{Stojchevska2014,Duan2021,Kogar2020}, but how defects are formed at the nanometer lengthscale and femtosecond timescale remains unknown. Here, we employ an intense laser pulse to create topological defects in a 2D charge density wave, and track their morphology and dynamics with ultrafast electron diffraction. Leveraging its high temporal resolution and sensitivity in detecting weak diffuse signals, we discover a dual-stage growth of 1D domain walls within 1~ps, a process not dictated by the order parameter amplitude but instead mediated by a nonthermal population of longitudinal optical phonons. Our work provides a framework for ultrafast engineering of topological defects based on selective excitation of collective modes, opening new avenues for dynamical control of nonequilibrium phases in correlated materials.}

An important pathway to realize emergent state out of equilibrium is via the creation of topological defects. These singularities not only modify the local amplitude and phase of the underlying order parameter, they also change the elementary excitation of a symmetry-broken state \cite{Yusupov2010,Zong2018}. In strongly correlated materials, a defect-induced renormalization of the electronic structure has been shown to lead to dramatic outcomes such as photoinduced insulator-metal transitions \cite{Stojchevska2014} and superconducting-like behavior \cite{Duan2021}. In systems that host several proximal states of matter, these defects can further seed the transient growth of a competing phase that is normally hidden in equilibrium \cite{Kogar2020}. Despite the ubiquity of nonequilibrium topological defects, their ephemeral nature makes it challenging to obtain a precise characterization of their evolution in space and time. Most experiments so far have been restricted to studying their annihilation dynamics from a few picoseconds to several minutes \cite{Vogelgesang2018,Zong2019,Laulhe2017,Chuang1991,Bowick1994}, but it remains unclear how individual defects are formed in the first place at a much faster timescale.

\begin{figure*}[htb!]
	\includegraphics[scale=0.54]{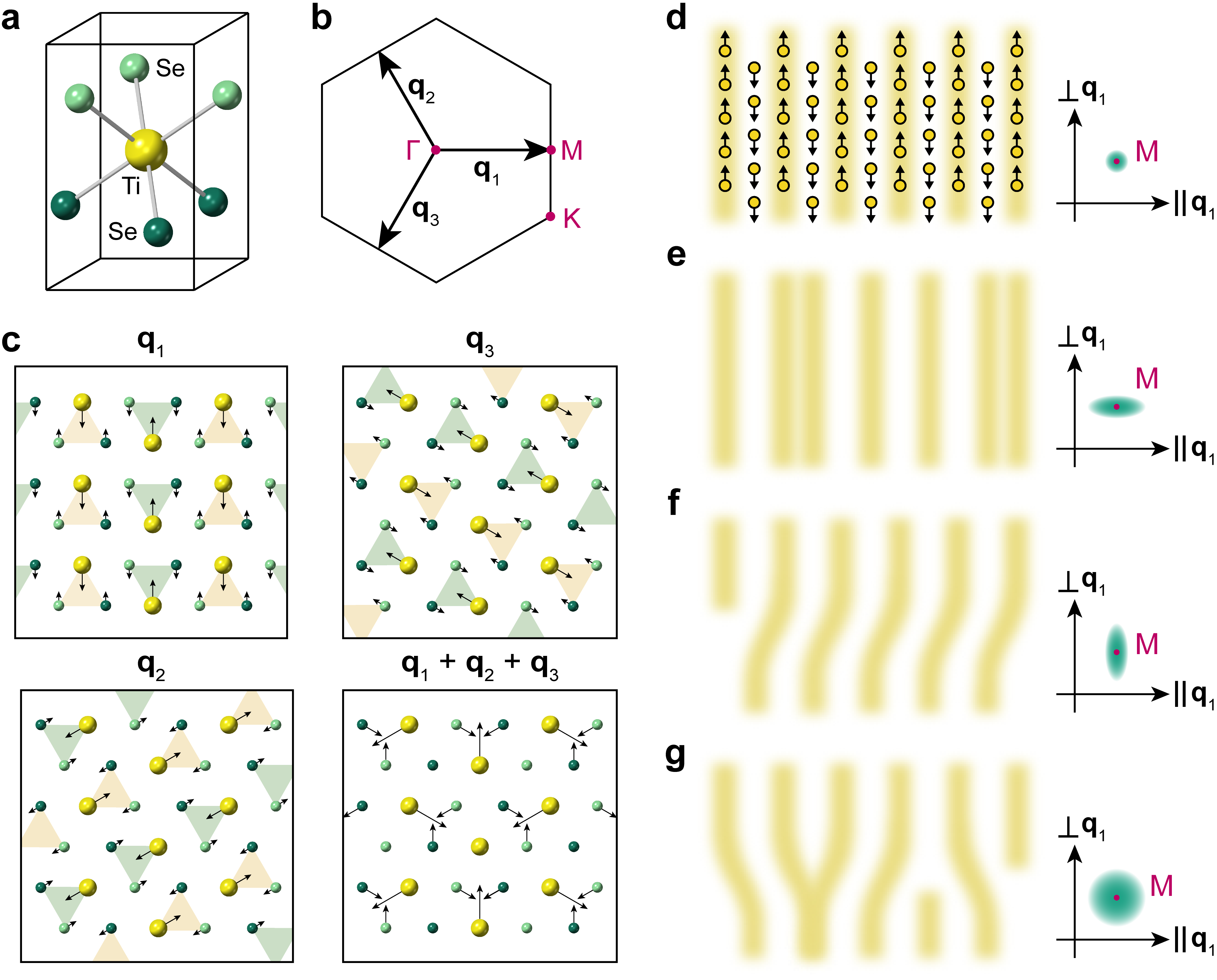}
	\caption{\textbf{$\bm{2\times2}$ charge density wave (CDW) in 1$\bm{T}$-TiSe$_\text{2}$ and possible types of topological defects in the CDW.} \textbf{a},~Crystal unit cell of 1$T$-TiSe$_2$ in the non-CDW state, where each Ti atom is octahedrally coordinated with six Se atoms. The upper and lower Se atoms are colored differently. \textbf{b},~Projected Brillouin zone of the non-CDW state with high symmetry points labeled. Arrows indicate the three wavevectors of the short-range $2\times2$ CDW, which exists right above the transition temperature of the long-range $2\times2\times2$ CDW order. \textbf{c},~Top view of the crystal structure, where CDW-induced atomic displacements in a single layer are labeled by arrows for the indicated wavevector ($\mathbf{q}_1$ to $\mathbf{q}_3$). For visual clarity, the length of displacement vectors and the ratio between the Ti and Se displacements are not drawn to scale. Green and beige triangles highlight alternating chains of anti-phase displacements in each wavevector. \textbf{d}--\textbf{g},~Schematics of different topological defects in the $2\times2$ CDW for wavevector $\mathbf{q}_1$. Right panels show the corresponding superlattice peak shape at the M point imaged in a 2D detector. Panel~\textbf{d} shows the state free from defects, where only Ti displacements are drawn. Yellow and white stripes correspond to CDW displacement chains with opposite phases. Possible topological defects include 1D domain walls~(\textbf{e}), shear~(\textbf{f}), and dislocations~(\textbf{g}).}
\label{fig:1}
\end{figure*}

Two-dimensional (2D) materials have been of great interest in the investigation of defect dynamics since pioneering studies on 2D liquid crystals \cite{Orihara1986,Nagaya1992,Nakai1996}. At equilibrium, topological defects readily appear due to pronounced thermal fluctuations in a reduced dimension, giving rise to exotic phases involving nematic, tetratic, or hexatic orders depending on the underlying crystal symmetry \cite{Nelson1979,Dai1991}. While scanning microscopes can image individual defects down to the atomic scale \cite{Soumyanarayanan2013,Ma2016,Cho2016}, dynamics of nanoscopic defects are often too fast to capture with traditional probes. To understand the mechanism of defect formation at the femtosecond timescale, we use an ultrashort light pulse to create topological defects in a 2D charge density wave (CDW), and we examine their temporal evolution encoded in diffuse scatterings in momentum space \cite{Chaikin1995}. Unlike specular peaks, these diffuse intensities are orders-of-magnitude smaller, rendering quantitative analysis difficult. We developed a femtosecond MeV electron diffraction beamline, whose high electron flux enabled an ultra-sensitive probe of diffuse signals without compromising the temporal resolution \cite{Qi2020} (see Methods). From the same set of diffraction images, we were able to simultaneously record branch-specific phonon dynamics \cite{Stern2018} as topological defects were formed. The concurrent detection of both topological defects and nonthermal phonons offers a striking visualization of different stages in defect growth, providing an unprecedented view on the mechanism of ultrafast defect generation in solids.

Our material choice is a layered transition metal dichalcogenide, 1$T$-TiSe$_2$ (Fig.~\ref{fig:1}a), which forms a $2\times2\times2$ commensurate CDW below $T_c\approx195$~K. In each layer, the density modulation is a superposition of three symmetry-equivalent wavevectors, $\mathbf{q}_1$ to $\mathbf{q}_3$ (Fig.~\ref{fig:1}b and \ref{fig:1}c), whose associated atomic displacements in adjacent layers are exactly $\pi$ out of phase. Right above $T_c$, photoemission and diffraction measurements reveal a 2D version of the ground state CDW \cite{Chen2016,Cheng2022}, featuring short-range $2\times2$ superlattices in individual layers that are largely uncoupled between layers. The existence of this 2D CDW allows us to explore defect dynamics in a reduced dimension without resorting to an atomically thin crystal.

\begin{figure*}[htb!]
\centering
    \includegraphics[width=0.95\textwidth]{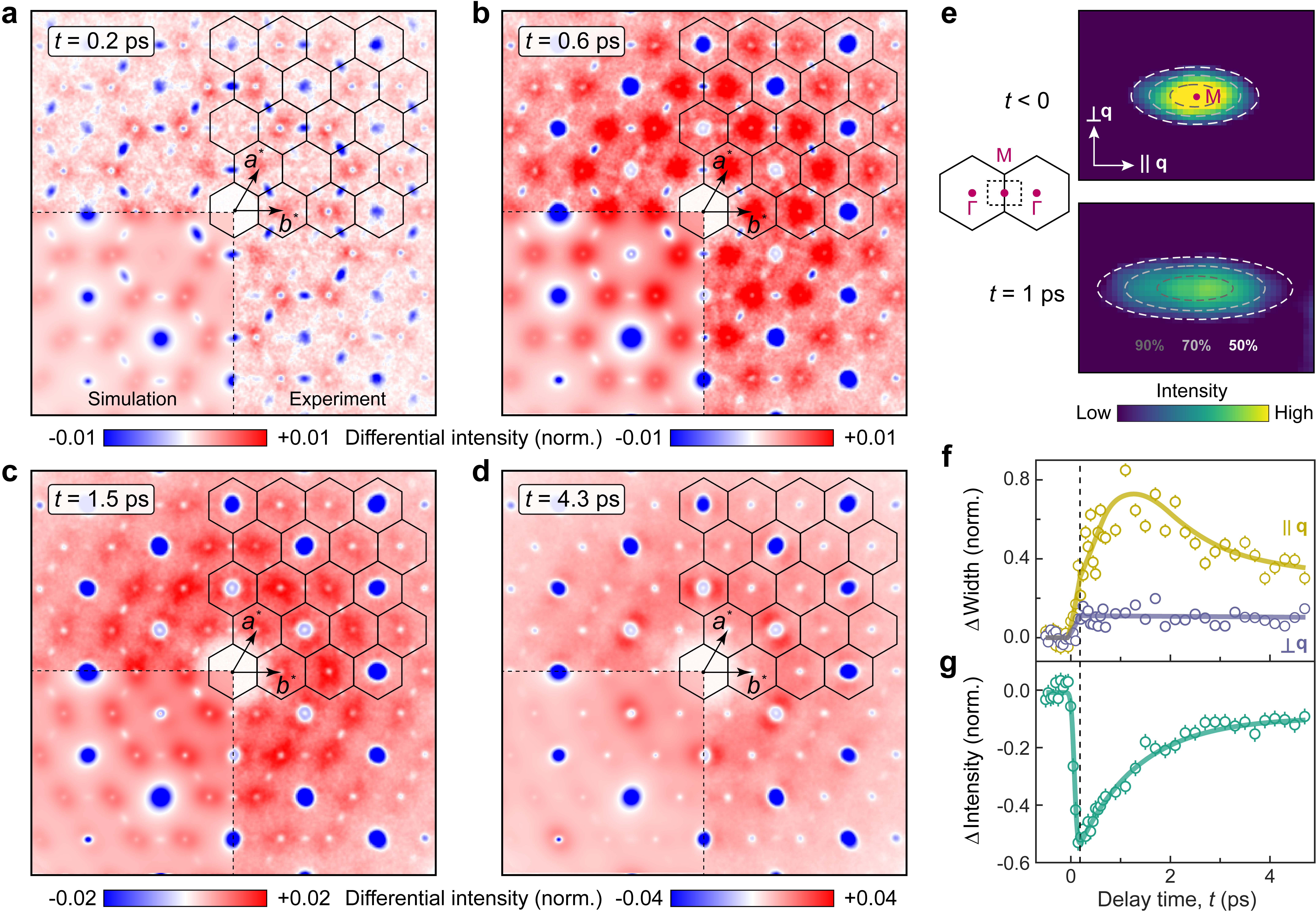}
    \caption{\textbf{Photoinduced 1D domain walls in the CDW.} \textbf{a}--\textbf{d},~Differential electron diffraction intensity at four time delays after photoexcitation by a 3-mJ/cm$^2$, 800-nm pulse; data were taken at 210~K. The lower left quadrant shows the simulation result \cite{SM}. Intensity changes are normalized by the pre-excitation values; note the different colorscales in different panels. Hexagons mark the Brillouin zones and images are symmetrized to enhance statistics. See Fig.~S2a--d for unsymmetrized patterns and Supplementary Video~1 for a continuous evolution of the diffraction intensity. \textbf{e},~Comparison of the 2D CDW peak at the M point before and 1~ps after photoexcitation, where signals from multiple M points that are symmetry-equivalent to $(H,K)=(3,-1.5)$ are averaged. Dashed rectangle in the inset marks the bounding box of the images. Dashed ellipses are fitted intensity contours at different levels relative to the peak intensity. \textbf{f},~Time evolution of the change in peak width at M. Significantly more broadening is observed in the direction parallel to the CDW wavevector ($\parallel\mathbf{q}$) than the perpendicular direction ($\perp\mathbf{q}$). Solid curves are guides to the eye. \textbf{g},~Time evolution of the change in peak intensity at M. Vertical dashed line marks the minimum of the transient intensity. The solid curve is a fit to an error function multiplied by an exponential. In \textbf{f} and \textbf{g}, all quantities are normalized to their respective values prior to pump incidence; error bars represent the standard deviation of their pre-excitation values.}
\label{fig:2}
\end{figure*}

In diffraction experiments, the type of CDW defects can be classified by the spatial profile of the superlattice peak. Focusing on a particular wavevector, such as $\mathbf{q}_1$, we expect a $\delta$-function-like peak at the M point if no defects are present, corresponding to perfect atomic displacements that form anti-phase chains running perpendicular to $\mathbf{q}_1$ (Fig.~\ref{fig:1}d). If defects are present, the $\delta$ peak gains a finite width \cite{Remark1}, whose anisotropy is indicative of the defect structure. In the context of a stripe-like CDW, an isotropic peak broadening points to the formation of dislocations (Fig.~\ref{fig:1}g) \cite{SM}. If broadening is predominantly perpendicular to $\mathbf{q}_1$, coherence between neighboring displacement chains are maintained but are reduced along the chain direction, consistent with shear (Fig.~\ref{fig:1}f). On the other hand, peak broadening parallel to $\mathbf{q}_1$ signifies the appearance of 1D domain walls, which consist of adjacent chains with in-phase atomic displacements (Fig.~\ref{fig:1}e). In all cases, the CDW defects are topological and they can only be created or annihilated in defect-antidefect pairs\cite{Chaikin1995}. 

The experiments were carried out on a freestanding thin flake of 1$T$-TiSe$_2$ at 210~K in its 2D CDW state (see Methods). Following photoexcitation by a 3-mJ/cm$^2$, 800-nm pulse, the change of diffraction pattern recorded with MeV relativistic electrons is shown in Fig.~\ref{fig:2}a--d along the [001] zone axis; see Supplementary Video~1 for a continuous evolution of the diffraction intensity. Thanks to the high electron energy, these differential maps reveal a rich set of diffuse patterns with a wide momentum range that are less affected by dynamical scattering effects compared to keV electrons \cite{Zhu2015}, yielding critical information about the microscopic environment surrounding the creation of CDW defects.

We first focus on photoinduced changes near the M point that report the 2D CDW dynamics. The emergence of blue spots in the difference map shown in Fig.~\ref{fig:2}a indicates a transient suppression of the CDW amplitude \cite{Cheng2022,Otto2021}. Their elongated shape with the long axis oriented perpendicular to the Brillouin zone boundary further suggests anisotropic peak broadening. In Fig.~\ref{fig:2}e, we zoom into a particular M point corresponding to one of the three CDW wavevectors, and we examine the diffuse intensity profile before and 1~ps after photoexcitation. As illustrated in the fitted contours (dashed ellipses), the peak width significantly increases in the direction parallel to the CDW wavevector ($\parallel\mathbf{q}$) while much smaller broadening occurs in the perpendicular direction ($\perp\mathbf{q}$). As the CDW correlation length is inversely proportional to the peak width, we estimate the correlation length to be 5.3~unit cells (defined in the non-CDW state) parallel to $\mathbf{q}$ and 17.3~unit cells perpendicular to $\mathbf{q}$ at 1~ps \cite{SM}. Based on the schematic in Fig.~\ref{fig:1}e, we therefore conclude that photoexcitation primarily gives rise to 1D domain walls while preserving the correlation along individual displacement chains. To confirm our defect assignment, we simulated the change of the M point intensity using anisotropic CDW correlations for all three wavevectors \cite{SM}. The computed patterns displayed in the bottom left quadrant of Fig.~\ref{fig:2}a--d demonstrate an excellent agreement with the experimental data.

These chain-like domain walls distinguish 1$T$-TiSe$_2$ from previous experimental and theoretical studies of transient defects in CDW compounds where dislocation pairs are created by photoexcitation \cite{Vogelgesang2018,Zong2019,Domrose2022,Tarkhov2022}. Here, the preferential formation of 1D domain walls instead of point-like dislocations validates an early prediction that CDWs in 1$T$-TiSe$_2$ can be viewed as weakly linked chains \cite{VanWezel2010}. From the theoretical standpoint, both second-order Jahn-Teller distortion and excitonic interaction between Se 4$p$ holes and Ti 3$d$ electrons are thought to account for the underlying 1D nature of the CDW in equilibrium. In our experiments, the high photoexcitation density leads to transient suppression of excitonic correlations due to screening by excited free carriers \cite{Porer2014}, suggesting that excitons do not play a significant role in maintaining this 1D chain configuration in the nonequilibrium context.

To understand how the 1D domain walls are created in the 2D CDW, in Fig.~\ref{fig:2}f, we plot the M point peak widths in two orthogonal directions as a function of pump-probe delay. For comparison, we also show the time evolution of the peak intensity in Fig.~\ref{fig:2}g, which reports the local amplitude of the order parameter \cite{Zong2021}. Within $200$~fs after photoexcitation, the CDW amplitude reaches a minimum, as indicated by the vertical dashed line. A similar timescale (200--300~fs) marks the initial peak broadening in both directions: parallel (yellow circles) and perpendicular (blue circles) to the CDW wavevector $\mathbf{q}$. After 300~fs, the CDW amplitude partially recovers and the peak width perpendicular to $\mathbf{q}$ remains constant. On the other hand, peak width parallel to $\mathbf{q}$ continues to grow significantly up to $\sim1$~ps, leading to an increasingly more anisotropic peak. These observations allow us to divide the domain wall formation into two stages. First, the transient loss of local CDW amplitude triggers a concurrent suppression of the CDW coherence, suggesting that the atomic displacement towards the high symmetry position is not executed in a concerted manner. Instead, significant disordering occurs in the atomic movement \cite{Wall2018}, especially between different CDW chains. In the second stage, further development of chain-like domain walls relies on a partial CDW amplitude recovery, where adjacent chains adopt an incorrect phase as atoms move back to the distorted lattice positions. Beyond $\sim1$~ps, domain walls start the slow annihilation process, and the change in the M point peak width remains highly anisotropic throughout our time delay window up to 4.7~ps.

\begin{figure*}[htb!]
    \includegraphics[width=0.95\textwidth]{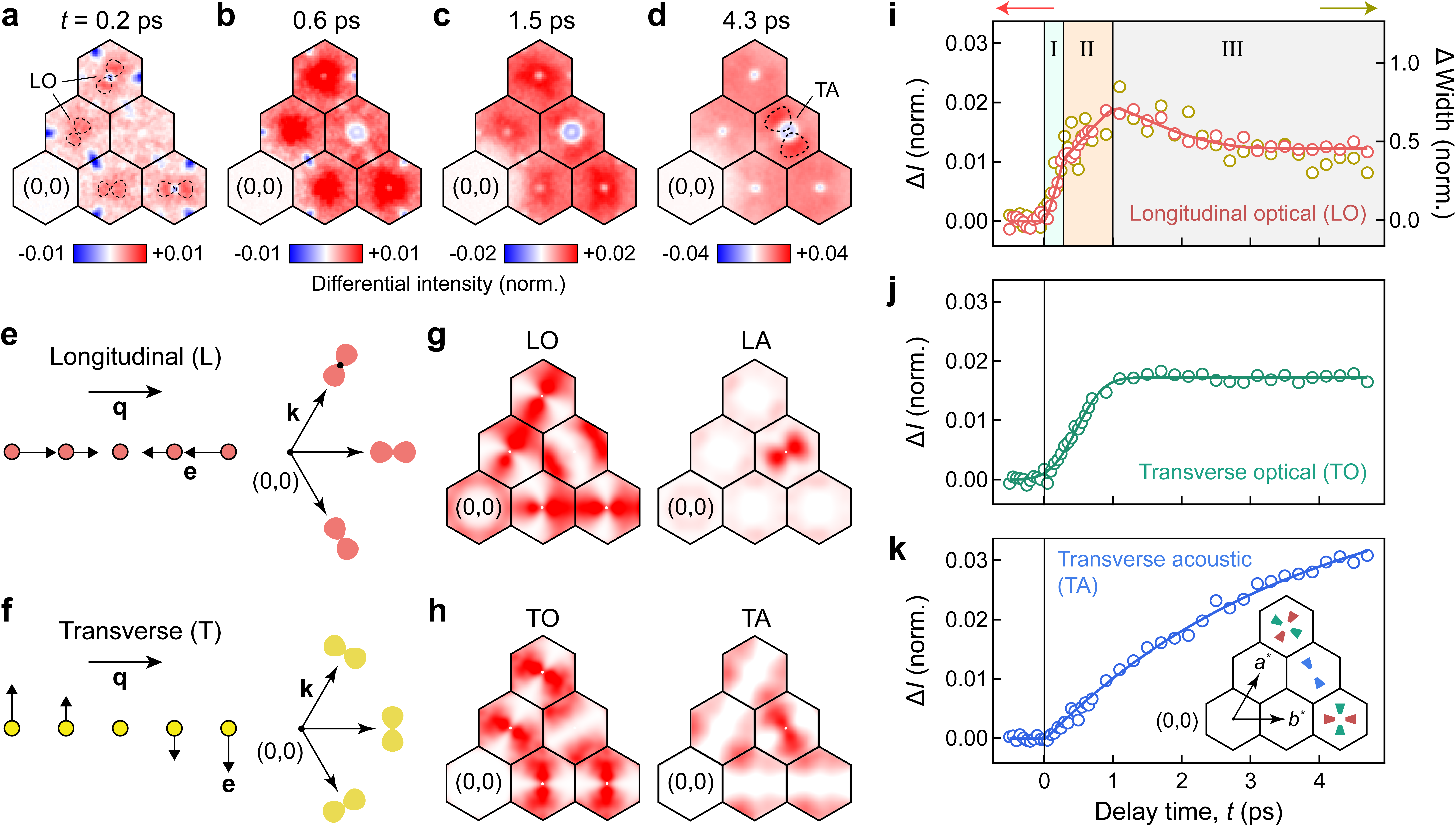}
    \caption{\textbf{Photoinduced phonon dynamics in different branches.} \textbf{a}--\textbf{d},~Differential diffraction patterns taken from Fig.~\ref{fig:2}a--d, zooming into the six Brillouin zones near the undiffracted beam at $(0,0)$. Regions of interest marked by dashed curves in \textbf{a} and \textbf{d} highlight the diffuse scatterings from longitudinal optical phonons (LO) and transverse acoustic phonons (TA), respectively. \textbf{e},\textbf{f},~Schematics of atomic displacements (left) and diffuse scatterings (right) for longitudinal (\textbf{e}) and transverse (\textbf{f}) phonons for a given wavevector $\mathbf{q}$, which is defined from the closest $\Gamma$ point. The scattering wavevector $\mathbf{k}$ is measured with respect to $(0,0)$. \textbf{g},\textbf{h},~Calculated one-phonon structure factor for longitudinal (L) and transverse (T) phonons from optical (O) and acoustic (A) branches. The structure factor of optical phonons concentrates in Brillouin zones with $H-K\neq0~(\text{mod}~3)$; the opposite holds for acoustic phonons. See Figs.~S4 and S5 for full images with more Brillouin zones included. \textbf{i}--\textbf{k},~Time evolution of diffuse intensities for different phonon branches, taken from color-coded regions of interests in the inset of panel~\textbf{k}. Intensities from symmetry-equivalent Brillouin zones are averaged. Solid curves are guides to the eye. In panel~\textbf{i}, the time trace of the 2D CDW peak width parallel to the CDW wavevector is overlaid (yellow circles, reproduced from Fig.~\ref{fig:2}f). Both the width and LO phonon population show similar three-part dynamics, labeled I--III, which are qualitatively distinct from TO and TA phonons in panels \textbf{j} and \textbf{k}.}
\label{fig:3}
\end{figure*}

While we have identified two stages of defect formation separated by the transient minimum of the CDW amplitude, it remains elusive what microscopic process contributes to phase disordering between adjacent chains during either stage. As domain wall formation fundamentally represents a special type of lattice dynamics, we look into nonthermal phonon populations induced by the femtosecond laser pulse, whose evolution is encoded in diffuse scattering signals away from Bragg peaks ($\Gamma$ point) and 2D CDW peaks (M point). In Fig.~\ref{fig:3}a--d, we re-examine the differential intensity patterns, zooming into the six Brillouin zones closest to the $(0,0)$ order. Based on the momentum anisotropy of red diffuse features, we can categorize them into either \textit{longitudinal} or \textit{transverse} phonons, shown schematically in Fig.~\ref{fig:3}e and \ref{fig:3}f. For example, at 0.2~ps, elongated diffuse signals resembling an hourglass shape are observed near Brillouin zone centers (dashed curves in Fig.~\ref{fig:3}a), where the elongation is parallel to the line connecting the zone center ($\Gamma$ point) and $(0,0)$. These are primarily longitudinal phonons, whose polarization vector $\mathbf{e}$ is parallel to the phonon wavevector $\mathbf{q}$ defined relative to the $\Gamma$ point. This assignment originates from the fact that the one-phonon structure factor contains a term $(\mathbf{e}\cdot\mathbf{k})$, where $\mathbf{k}$ is the scattering wavevector measured from $(0,0)$ \cite{Xu2005}. Hence, the most prominent diffuse intensity appears in regions of the Brillouin zone where $\mathbf{q}\parallel\mathbf{k}$. The same argument implies that transverse phonons near the $\Gamma$ point appear in similarly shaped diffuse features but are 90$^\circ$ rotated (Fig.~\ref{fig:3}f). This is the case at 4.3~ps in the $(1,1)$ Brillouin zone (dashed curves in Fig.~\ref{fig:3}d). Using phonon structure factors obtained from first-principles calculations, we can further break down longitudinal and transverse phonons into \textit{optical} and \textit{acoustic} branches \cite{SM}. Specifically, diffuse signals from optical phonons are more intense in $(H,K)$ zones where $(H-K)\neq 0~(\text{mod}~3)$; the opposite applies to acoustic phonons (Fig.~\ref{fig:3}g and \ref{fig:3}h). Hence, at 0.2~ps, the longitudinal optical phonons are the first to be excited among all branches (Fig.~\ref{fig:3}a) while the quasi-equilibrium state at 4.3~ps is characterized by dominant population of transverse acoustic phonons (Fig.~\ref{fig:3}d). These assignments are further confirmed by diffuse scattering simulations in Fig.~\ref{fig:2}a--d (bottom left quadrant), which accurately reproduce the experimental observation, including the suppression of Bragg peaks at long time delay primarily due to the acoustic phonons emitted.

To quantify the nonthermal phonon population, we plot the diffuse intensity evolutions in Fig.~\ref{fig:3}i--k for the three dominant phonons observed: longitudinal optical (LO), transverse optical (TO), and transverse acoustic (TA). Below 300~fs, the LO phonons are preferentially excited, which continue to grow after 300~fs but at a smaller rate of increase (Fig.~\ref{fig:3}i). Their population subsequently decreases after 1~ps and reaches a metastable value. This three-part evolution of the LO phonons (labeled~I--III in Fig.~\ref{fig:3}i) distinguishes them from TO and TA phonons, which only feature a rise on slower timescales without any relaxation over a few picoseconds (Fig.~\ref{fig:3}j and \ref{fig:3}k). Notably, the LO phonon dynamics (red circles in Fig.~\ref{fig:3}i) almost exactly coincide with the change in the CDW peak width in the direction parallel to the CDW wavevector (yellow circles in Fig.~\ref{fig:3}i), both showing a dual-stage growth followed by a relaxation with identical timescales. As the peak width scales with the domain wall density, this correlation suggests that the emission of LO phonons from excited electrons play a critical role in creating the topological defects in the 2D CDW.

\begin{figure*}[htb!]
    \includegraphics[scale=0.5]{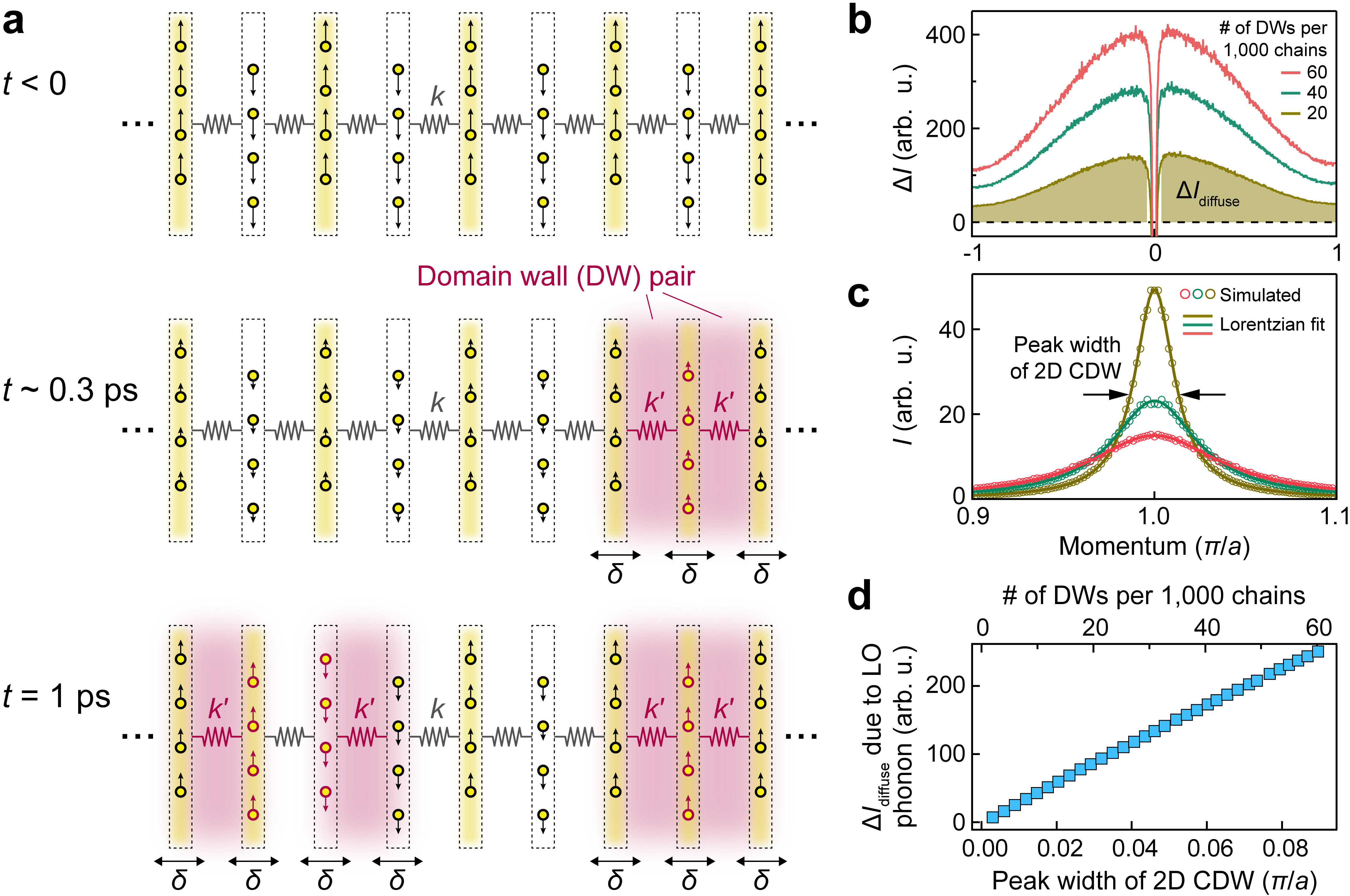}
    \caption{\textbf{Ultrafast domain wall creation mediated by longitudinal optical phonons.} \textbf{a},~Schematic of dual-stage formation of CDW domain walls illustrated for a particular CDW wavevector. Only Ti atoms are shown (circles), where arrows denote CDW displacements. In the first stage (up to $\sim0.3$~ps), the CDW amplitude decreases, represented by the shrinking arrow length; in the second stage (up to 1~ps), the CDW amplitude partially recovers. Pairs of domain walls are indicated by red shades, where CDW displacements in the opposite direction compared to the $t<0$ configuration are highlighted by red arrows. Domain wall formation modifies the local inter-chain coupling ($k\rightarrow k'$), leading to local displacements ($\delta$) that constitute the longitudinal optical phonons. \textbf{b},~Simulated change of diffuse scattering intensity due to longitudinal phonons arising from local atomic displacements at domain walls ($\delta$ motion in panel~\textbf{a}). Simulations were performed for a variable numbers of domain wall (DW) densities. The dip at the zero momentum (the $\Gamma$ point) reflects the Debye-Waller intensity drop of the Bragg peak. The shaded area represents the momentum range for calculating the integrated intensity $\Delta I_\text{diffuse}$ used in panel \textbf{d}. \textbf{c},~Simulated CDW peak at the Brillouin zone boundary for various domain wall densities (same color coding as in panel~\textbf{b}). Solid curves are Lorentzian fits. \textbf{d},~Simulated diffuse intensity change arising from LO phonons as a function of CDW domain wall densities (\textit{top axis}), or equivalently, the width of the 2D CDW peak (\textit{bottom axis}).}
\label{fig:4}
\end{figure*}

We can gain some insight into the coupling between the LO phonons and 1D domain walls by considering the model of linked CDW chains for each of the three CDW wavevectors \cite{VanWezel2010,SM}. Here, we treat each chain as an individual entity coupled by a spring constant $k$, where LO phonons manifest as local, non-propagating oscillations that modulate the inter-chain distance (Fig.~\ref{fig:4}a). The most drastic change in inter-chain coupling is expected to occur at CDW domain walls due to the in-phase atomic displacements in two neighboring chains (shaded regions in red). This modification of the spring constant from $k$ to $k'$ inevitably leads to a change of the local inter-chain distance and hence the excitation of LO phonons at the domain wall ($\delta$ vibrations in Fig.~\ref{fig:4}a). Conversely, nonthermal population of LO phonons can result in large inter-chain displacements, where in-phase rather than out-of-phase CDW distortions can be favored locally, seeding the growth of a domain wall. To be more quantitative, we compute from this model the change in diffuse intensity associated with LO phonons ($\Delta I_\text{diffuse}$, Fig.~\ref{fig:4}b) for various domain wall densities, which are proportional to the peak widths of the 2D CDW in the direction parallel to the CDW wavevector (Fig.~\ref{fig:4}c). The nearly perfect linear relation between $\Delta I_\text{diffuse}$ and the peak width shown in Fig.~\ref{fig:4}d provides a natural explanation for their overlapping temporal evolutions observed in Fig.~\ref{fig:3}i. Our calculation further predicts that the momentum-space profile of the phonon diffuse signal remains unaltered despite a change in the intensity (Fig.~S9e). This universal line profile is clearly evidenced in the experimental diffraction pattern, lending further support to our model \cite{SM}. 

The simultaneous measurements of both 2D CDW peaks and diffuse phonon signals offer a detailed picture of how topological defects develop in two stages in the sub-picosecond regime, as summarized in Fig.~\ref{fig:4}a. Within the first $\sim300$~fs --- a timescale not limited by the instrument temporal resolution (see Methods) --- a high density of free carriers excited by the laser pulse transiently suppresses the CDW amplitude. At the same time, a rapidly increasing LO phonon population significantly modulates the inter-chain interaction, leading to predominantly local flips of the CDW displacement direction and the production of pairs of 1D domain walls. During the second stage that lasts up to 1~ps, the CDW amplitude starts to recover, but continued emission of the LO phonons creates local minima in the energy landscape where in-phase CDW displacements between neighboring chains are preferred. Hence, more pairs of domain walls are formed, which are not annihilated until several picoseconds or longer. 

The 1D domain walls identified in this study disrupts the translational order of the $2\times2$ CDW in 1$T$-TiSe$_2$ while respecting the threefold rotational symmetry. The resulting state is consistent with the hexatic order that appears as an intermediate phase when a 2D triangular lattice is melted by dissociation of bound topological defects. Our work provides the missing time axis in the sub-picosecond scale for describing this defect-driven transition, highlighting the role of longitudinal phonons in mediating the domain wall generation. We expect the mechanism discovered here to be general for other symmetry-broken states, where defect generation takes place regardless of the order parameter amplitude but is instead associated with nonthermal excitation of specific collective modes. This result paves the way for ultrafast manipulation of topological defects in different ground states, a fruitful avenue for discovering new phases of matter in quantum materials and beyond \cite{DelaTorre2021}.

\section{Methods}

\noindent\textbf{Sample growth and preparation.}~High-quality single crystals of 1$T$-TiSe$_2$ were grown by chemical vapour transport with an iodine transport agent. Ti and Se were mixed in a molar ratio of 1:2 and placed into an alumina crucible before being sealed into a quartz tube. The quartz tube was heated to 700$^\circ$C and 1$T$-TiSe$_2$ crystals were synthesized at the 650$^\circ$C zone for two weeks. 1$T$-TiSe$_2$ thin flakes were obtained by repeated exfoliation of the bulk crystal with polydimethylsiloxane films (PDMS, Gel-Pak). Flakes were pre-screened for thickness and uniformity with an optical microscope using the color contrast and further characterized by atomic force microscopy. Selected flakes were detached from PDMS in ethanol and scooped onto standard copper TEM grids. The resulting free-standing flake has a typical lateral dimension of $\sim400~\upmu$m and thickness of $\sim30$~nm. A typical optical image of a flake is shown in Fig.~S10a.\\

\noindent\textbf{MeV ultrafast electron diffraction.}~Details of the ultrafast electron diffraction beamline are discussed in ref.~\cite{Qi2020}. Briefly, the 800-nm (1.55-eV), 30-fs pulses from a Ti:sapphire regenerative amplifier system operating at a repetition rate of 100~Hz (Vitara and Legend Elite Duo HE, Coherent) were split into pump and probe branches. The probe branch was frequency tripled in nonlinear crystals before illuminating a photocathode for electron pulse generation, producing approximately 30,000 electrons per pulse (5~fC). After being accelerated by an intense radio-frequency field to relativistic velocity ($\sim 0.989c$), the electron beam went through a double-bend achromatic lens for pulse compression and jitter removal. The typical electron beam spot size on the sample was approximately 150~$\upmu$m measured at full-width at half maximum (FWHM), nearly five times smaller than the size of the pump pulse, ensuring a homogeneous photoexcitation condition. The temporal delay between the pump and probe pulses was adjusted by a linear translation stage, and the temporal resolution was 50~fs. Diffracted electron beams were incident on a phosphor screen (P43) and the image was collected by an electron-multiplying charge-coupled device. The sample was cooled by liquid nitrogen.

\section{Additional Information}

\noindent\textbf{Acknowledgements.}~
We thank Yu~He and David~Limmer for helpful discussions. 
We thank Jan-Hendrik P\"ohls for providing the calculated phonon dispersions in ref.~\cite{Otto2021} and Laurent P. Ren\'e de Cotret for providing the Python codes for computing the one-phonon structure factor in ref.~\cite{Stern2018}. 
D.X. and J.Z. acknowledge support from the National Key R\&D Program of China (No.~2021YFA1400202), the National Natural Science Foundation of China (grant No.'s~11925505, 11504232 and 11721091), and the office of Science and Technology, Shanghai Municipal Government (No.~16DZ2260200). 
A.Z. acknowledges support from the Miller Institute for Basic Research in Science. 
L.W., Q.M., and Y.Z. acknowledge support from the U.S. Department of Energy, Basic Energy Sciences, Materials Sciences and Engineering Division under Contract No. DE-SC0012704. 
Y.G. acknowledges support from the National Natural Science Foundation of China (grant No.~11874264). 
M.W.Z. acknowledges funding by the W.~M. Keck Foundation, funding from the UC Oﬃce of the President within the Multicampus Research Programs and Initiatives (M21PL3263), and the Hellman Fellows Fund. \\

\noindent\textbf{Author contributions.}~
Y.C., D.X., and A.Z. designed the research. Y.C., A.Z., L.W., and Y.Z. performed data analysis with important theoretical insights from J.V.W. and A.K. 
W.X. and Y.G. grew the single crystals. 
Y.C. collected the UED data, where the MeV-UED beamline was constructed and maintained by F.Q., P.Z., X.Z., and T.J. 
L.W., Q.M., and Y.C. performed diffraction simulations.  
A.Z. and Y.C. wrote the manuscript with critical input from A.K., M.W.Z., Y.Z., D.X., and all other authors. 
The project was supervised by M.W.Z., J.Z., Y.Z., and D.X.


\begin{thebibliography}{38}%
\makeatletter
\providecommand \@ifxundefined [1]{%
 \@ifx{#1\undefined}
}%
\providecommand \@ifnum [1]{%
 \ifnum #1\expandafter \@firstoftwo
 \else \expandafter \@secondoftwo
 \fi
}%
\providecommand \@ifx [1]{%
 \ifx #1\expandafter \@firstoftwo
 \else \expandafter \@secondoftwo
 \fi
}%
\providecommand \natexlab [1]{#1}%
\providecommand \enquote  [1]{``#1''}%
\providecommand \bibnamefont  [1]{#1}%
\providecommand \bibfnamefont [1]{#1}%
\providecommand \citenamefont [1]{#1}%
\providecommand \href@noop [0]{\@secondoftwo}%
\providecommand \href [0]{\begingroup \@sanitize@url \@href}%
\providecommand \@href[1]{\@@startlink{#1}\@@href}%
\providecommand \@@href[1]{\endgroup#1\@@endlink}%
\providecommand \@sanitize@url [0]{\catcode `\\12\catcode `\$12\catcode
  `\&12\catcode `\#12\catcode `\^12\catcode `\_12\catcode `\%12\relax}%
\providecommand \@@startlink[1]{}%
\providecommand \@@endlink[0]{}%
\providecommand \url  [0]{\begingroup\@sanitize@url \@url }%
\providecommand \@url [1]{\endgroup\@href {#1}{\urlprefix }}%
\providecommand \urlprefix  [0]{URL }%
\providecommand \Eprint [0]{\href }%
\providecommand \doibase [0]{https://doi.org/}%
\providecommand \selectlanguage [0]{\@gobble}%
\providecommand \bibinfo  [0]{\@secondoftwo}%
\providecommand \bibfield  [0]{\@secondoftwo}%
\providecommand \translation [1]{[#1]}%
\providecommand \BibitemOpen [0]{}%
\providecommand \bibitemStop [0]{}%
\providecommand \bibitemNoStop [0]{.\EOS\space}%
\providecommand \EOS [0]{\spacefactor3000\relax}%
\providecommand \BibitemShut  [1]{\csname bibitem#1\endcsname}%
\let\auto@bib@innerbib\@empty
\bibitem [{\citenamefont {Zurek}(1985)}]{Zurek1985}%
  \BibitemOpen
  \bibfield  {author} {\bibinfo {author} {\bibfnamefont {W.~H.}\ \bibnamefont
  {Zurek}},\ }\bibfield  {title} {\bibinfo {title} {{Cosmological experiments
  in superfluid helium?}},\ }\href {https://doi.org/10.1038/317505a0}
  {\bibfield  {journal} {\bibinfo  {journal} {Nature}\ }\textbf {\bibinfo
  {volume} {317}},\ \bibinfo {pages} {505} (\bibinfo {year}
  {1985})}\BibitemShut {NoStop}%
\bibitem [{\citenamefont {Keesling}\ \emph {et~al.}(2019)\citenamefont
  {Keesling}, \citenamefont {Omran}, \citenamefont {Levine}, \citenamefont
  {Bernien}, \citenamefont {Pichler}, \citenamefont {Choi}, \citenamefont
  {Samajdar}, \citenamefont {Schwartz}, \citenamefont {Silvi}, \citenamefont
  {Sachdev}, \citenamefont {Zoller}, \citenamefont {Endres}, \citenamefont
  {Greiner}, \citenamefont {Vuleti{\'{c}}},\ and\ \citenamefont
  {Lukin}}]{Keesling2019}%
  \BibitemOpen
  \bibfield  {author} {\bibinfo {author} {\bibfnamefont {A.}~\bibnamefont
  {Keesling}}, \bibinfo {author} {\bibfnamefont {A.}~\bibnamefont {Omran}},
  \bibinfo {author} {\bibfnamefont {H.}~\bibnamefont {Levine}}, \bibinfo
  {author} {\bibfnamefont {H.}~\bibnamefont {Bernien}}, \bibinfo {author}
  {\bibfnamefont {H.}~\bibnamefont {Pichler}}, \bibinfo {author} {\bibfnamefont
  {S.}~\bibnamefont {Choi}}, \bibinfo {author} {\bibfnamefont {R.}~\bibnamefont
  {Samajdar}}, \bibinfo {author} {\bibfnamefont {S.}~\bibnamefont {Schwartz}},
  \bibinfo {author} {\bibfnamefont {P.}~\bibnamefont {Silvi}}, \bibinfo
  {author} {\bibfnamefont {S.}~\bibnamefont {Sachdev}}, \bibinfo {author}
  {\bibfnamefont {P.}~\bibnamefont {Zoller}}, \bibinfo {author} {\bibfnamefont
  {M.}~\bibnamefont {Endres}}, \bibinfo {author} {\bibfnamefont
  {M.}~\bibnamefont {Greiner}}, \bibinfo {author} {\bibfnamefont
  {V.}~\bibnamefont {Vuleti{\'{c}}}},\ and\ \bibinfo {author} {\bibfnamefont
  {M.~D.}\ \bibnamefont {Lukin}},\ }\bibfield  {title} {\bibinfo {title}
  {{Quantum Kibble–Zurek mechanism and critical dynamics on a programmable
  Rydberg simulator}},\ }\href {https://doi.org/10.1038/s41586-019-1070-1}
  {\bibfield  {journal} {\bibinfo  {journal} {Nature}\ }\textbf {\bibinfo
  {volume} {568}},\ \bibinfo {pages} {207} (\bibinfo {year}
  {2019})}\BibitemShut {NoStop}%
\bibitem [{\citenamefont {Stojchevska}\ \emph {et~al.}(2014)\citenamefont
  {Stojchevska}, \citenamefont {Vaskivskyi}, \citenamefont {Mertelj},
  \citenamefont {Kusar}, \citenamefont {Svetin}, \citenamefont {Brazovskii},\
  and\ \citenamefont {Mihailovic}}]{Stojchevska2014}%
  \BibitemOpen
  \bibfield  {author} {\bibinfo {author} {\bibfnamefont {L.}~\bibnamefont
  {Stojchevska}}, \bibinfo {author} {\bibfnamefont {I.}~\bibnamefont
  {Vaskivskyi}}, \bibinfo {author} {\bibfnamefont {T.}~\bibnamefont {Mertelj}},
  \bibinfo {author} {\bibfnamefont {P.}~\bibnamefont {Kusar}}, \bibinfo
  {author} {\bibfnamefont {D.}~\bibnamefont {Svetin}}, \bibinfo {author}
  {\bibfnamefont {S.}~\bibnamefont {Brazovskii}},\ and\ \bibinfo {author}
  {\bibfnamefont {D.}~\bibnamefont {Mihailovic}},\ }\bibfield  {title}
  {\bibinfo {title} {{Ultrafast switching to a stable hidden quantum state in
  an electronic crystal}},\ }\href {https://doi.org/10.1126/science.1241591}
  {\bibfield  {journal} {\bibinfo  {journal} {Science}\ }\textbf {\bibinfo
  {volume} {344}},\ \bibinfo {pages} {177} (\bibinfo {year}
  {2014})}\BibitemShut {NoStop}%
\bibitem [{\citenamefont {Duan}\ \emph {et~al.}(2021)\citenamefont {Duan},
  \citenamefont {Cheng}, \citenamefont {Xia}, \citenamefont {Yang},
  \citenamefont {Xu}, \citenamefont {Qi}, \citenamefont {Huang}, \citenamefont
  {Tang}, \citenamefont {Guo}, \citenamefont {Luo}, \citenamefont {Qian},
  \citenamefont {Xiang}, \citenamefont {Zhang},\ and\ \citenamefont
  {Zhang}}]{Duan2021}%
  \BibitemOpen
  \bibfield  {author} {\bibinfo {author} {\bibfnamefont {S.}~\bibnamefont
  {Duan}}, \bibinfo {author} {\bibfnamefont {Y.}~\bibnamefont {Cheng}},
  \bibinfo {author} {\bibfnamefont {W.}~\bibnamefont {Xia}}, \bibinfo {author}
  {\bibfnamefont {Y.}~\bibnamefont {Yang}}, \bibinfo {author} {\bibfnamefont
  {C.}~\bibnamefont {Xu}}, \bibinfo {author} {\bibfnamefont {F.}~\bibnamefont
  {Qi}}, \bibinfo {author} {\bibfnamefont {C.}~\bibnamefont {Huang}}, \bibinfo
  {author} {\bibfnamefont {T.}~\bibnamefont {Tang}}, \bibinfo {author}
  {\bibfnamefont {Y.}~\bibnamefont {Guo}}, \bibinfo {author} {\bibfnamefont
  {W.}~\bibnamefont {Luo}}, \bibinfo {author} {\bibfnamefont {D.}~\bibnamefont
  {Qian}}, \bibinfo {author} {\bibfnamefont {D.}~\bibnamefont {Xiang}},
  \bibinfo {author} {\bibfnamefont {J.}~\bibnamefont {Zhang}},\ and\ \bibinfo
  {author} {\bibfnamefont {W.}~\bibnamefont {Zhang}},\ }\bibfield  {title}
  {\bibinfo {title} {{Optical manipulation of electronic dimensionality in a
  quantum material}},\ }\href {https://doi.org/10.1038/s41586-021-03643-8}
  {\bibfield  {journal} {\bibinfo  {journal} {Nature}\ }\textbf {\bibinfo
  {volume} {595}},\ \bibinfo {pages} {239} (\bibinfo {year}
  {2021})}\BibitemShut {NoStop}%
\bibitem [{\citenamefont {Kogar}\ \emph {et~al.}(2020)\citenamefont {Kogar},
  \citenamefont {Zong}, \citenamefont {Dolgirev}, \citenamefont {Shen},
  \citenamefont {Straquadine}, \citenamefont {Bie}, \citenamefont {Wang},
  \citenamefont {Rohwer}, \citenamefont {Tung}, \citenamefont {Yang},
  \citenamefont {Li}, \citenamefont {Yang}, \citenamefont {Weathersby},
  \citenamefont {Park}, \citenamefont {Kozina}, \citenamefont {Sie},
  \citenamefont {Wen}, \citenamefont {Jarillo-Herrero}, \citenamefont {Fisher},
  \citenamefont {Wang},\ and\ \citenamefont {Gedik}}]{Kogar2020}%
  \BibitemOpen
  \bibfield  {author} {\bibinfo {author} {\bibfnamefont {A.}~\bibnamefont
  {Kogar}}, \bibinfo {author} {\bibfnamefont {A.}~\bibnamefont {Zong}},
  \bibinfo {author} {\bibfnamefont {P.~E.}\ \bibnamefont {Dolgirev}}, \bibinfo
  {author} {\bibfnamefont {X.}~\bibnamefont {Shen}}, \bibinfo {author}
  {\bibfnamefont {J.}~\bibnamefont {Straquadine}}, \bibinfo {author}
  {\bibfnamefont {Y.-Q.}\ \bibnamefont {Bie}}, \bibinfo {author} {\bibfnamefont
  {X.}~\bibnamefont {Wang}}, \bibinfo {author} {\bibfnamefont {T.}~\bibnamefont
  {Rohwer}}, \bibinfo {author} {\bibfnamefont {I.-C.}\ \bibnamefont {Tung}},
  \bibinfo {author} {\bibfnamefont {Y.}~\bibnamefont {Yang}}, \bibinfo {author}
  {\bibfnamefont {R.}~\bibnamefont {Li}}, \bibinfo {author} {\bibfnamefont
  {J.}~\bibnamefont {Yang}}, \bibinfo {author} {\bibfnamefont {S.}~\bibnamefont
  {Weathersby}}, \bibinfo {author} {\bibfnamefont {S.}~\bibnamefont {Park}},
  \bibinfo {author} {\bibfnamefont {M.~E.}\ \bibnamefont {Kozina}}, \bibinfo
  {author} {\bibfnamefont {E.~J.}\ \bibnamefont {Sie}}, \bibinfo {author}
  {\bibfnamefont {H.}~\bibnamefont {Wen}}, \bibinfo {author} {\bibfnamefont
  {P.}~\bibnamefont {Jarillo-Herrero}}, \bibinfo {author} {\bibfnamefont
  {I.~R.}\ \bibnamefont {Fisher}}, \bibinfo {author} {\bibfnamefont
  {X.}~\bibnamefont {Wang}},\ and\ \bibinfo {author} {\bibfnamefont
  {N.}~\bibnamefont {Gedik}},\ }\bibfield  {title} {\bibinfo {title}
  {{Light-induced charge density wave in LaTe$_3$}},\ }\href
  {https://doi.org/10.1038/s41567-019-0705-3} {\bibfield  {journal} {\bibinfo
  {journal} {Nat. Phys.}\ }\textbf {\bibinfo {volume} {16}},\ \bibinfo {pages}
  {159} (\bibinfo {year} {2020})}\BibitemShut {NoStop}%
\bibitem [{\citenamefont {Yusupov}\ \emph {et~al.}(2010)\citenamefont
  {Yusupov}, \citenamefont {Mertelj}, \citenamefont {Kabanov}, \citenamefont
  {Brazovskii}, \citenamefont {Kusar}, \citenamefont {Chu}, \citenamefont
  {Fisher},\ and\ \citenamefont {Mihailovic}}]{Yusupov2010}%
  \BibitemOpen
  \bibfield  {author} {\bibinfo {author} {\bibfnamefont {R.}~\bibnamefont
  {Yusupov}}, \bibinfo {author} {\bibfnamefont {T.}~\bibnamefont {Mertelj}},
  \bibinfo {author} {\bibfnamefont {V.~V.}\ \bibnamefont {Kabanov}}, \bibinfo
  {author} {\bibfnamefont {S.}~\bibnamefont {Brazovskii}}, \bibinfo {author}
  {\bibfnamefont {P.}~\bibnamefont {Kusar}}, \bibinfo {author} {\bibfnamefont
  {J.-H.}\ \bibnamefont {Chu}}, \bibinfo {author} {\bibfnamefont {I.~R.}\
  \bibnamefont {Fisher}},\ and\ \bibinfo {author} {\bibfnamefont
  {D.}~\bibnamefont {Mihailovic}},\ }\bibfield  {title} {\bibinfo {title}
  {{Coherent dynamics of macroscopic electronic order through a symmetry
  breaking transition}},\ }\href {https://doi.org/10.1038/nphys1738} {\bibfield
   {journal} {\bibinfo  {journal} {Nat. Phys.}\ }\textbf {\bibinfo {volume}
  {6}},\ \bibinfo {pages} {681} (\bibinfo {year} {2010})}\BibitemShut {NoStop}%
\bibitem [{\citenamefont {Zong}\ \emph {et~al.}(2018)\citenamefont {Zong},
  \citenamefont {Shen}, \citenamefont {Kogar}, \citenamefont {Ye},
  \citenamefont {Marks}, \citenamefont {Chowdhury}, \citenamefont {Rohwer},
  \citenamefont {Freelon}, \citenamefont {Weathersby}, \citenamefont {Li},
  \citenamefont {Yang}, \citenamefont {Checkelsky}, \citenamefont {Wang},\ and\
  \citenamefont {Gedik}}]{Zong2018}%
  \BibitemOpen
  \bibfield  {author} {\bibinfo {author} {\bibfnamefont {A.}~\bibnamefont
  {Zong}}, \bibinfo {author} {\bibfnamefont {X.}~\bibnamefont {Shen}}, \bibinfo
  {author} {\bibfnamefont {A.}~\bibnamefont {Kogar}}, \bibinfo {author}
  {\bibfnamefont {L.}~\bibnamefont {Ye}}, \bibinfo {author} {\bibfnamefont
  {C.}~\bibnamefont {Marks}}, \bibinfo {author} {\bibfnamefont
  {D.}~\bibnamefont {Chowdhury}}, \bibinfo {author} {\bibfnamefont
  {T.}~\bibnamefont {Rohwer}}, \bibinfo {author} {\bibfnamefont
  {B.}~\bibnamefont {Freelon}}, \bibinfo {author} {\bibfnamefont
  {S.}~\bibnamefont {Weathersby}}, \bibinfo {author} {\bibfnamefont
  {R.}~\bibnamefont {Li}}, \bibinfo {author} {\bibfnamefont {J.}~\bibnamefont
  {Yang}}, \bibinfo {author} {\bibfnamefont {J.}~\bibnamefont {Checkelsky}},
  \bibinfo {author} {\bibfnamefont {X.}~\bibnamefont {Wang}},\ and\ \bibinfo
  {author} {\bibfnamefont {N.}~\bibnamefont {Gedik}},\ }\bibfield  {title}
  {\bibinfo {title} {{Ultrafast manipulation of mirror domain walls in a charge
  density wave}},\ }\href {https://doi.org/10.1126/sciadv.aau5501} {\bibfield
  {journal} {\bibinfo  {journal} {Sci. Adv.}\ }\textbf {\bibinfo {volume}
  {4}},\ \bibinfo {pages} {eaau5501} (\bibinfo {year} {2018})}\BibitemShut
  {NoStop}%
\bibitem [{\citenamefont {Vogelgesang}\ \emph {et~al.}(2018)\citenamefont
  {Vogelgesang}, \citenamefont {Storeck}, \citenamefont {Horstmann},
  \citenamefont {Diekmann}, \citenamefont {Sivis}, \citenamefont {Schramm},
  \citenamefont {Rossnagel}, \citenamefont {Sch{\"a}fer},\ and\ \citenamefont
  {Ropers}}]{Vogelgesang2018}%
  \BibitemOpen
  \bibfield  {author} {\bibinfo {author} {\bibfnamefont {S.}~\bibnamefont
  {Vogelgesang}}, \bibinfo {author} {\bibfnamefont {G.}~\bibnamefont
  {Storeck}}, \bibinfo {author} {\bibfnamefont {J.~G.}\ \bibnamefont
  {Horstmann}}, \bibinfo {author} {\bibfnamefont {T.}~\bibnamefont {Diekmann}},
  \bibinfo {author} {\bibfnamefont {M.}~\bibnamefont {Sivis}}, \bibinfo
  {author} {\bibfnamefont {S.}~\bibnamefont {Schramm}}, \bibinfo {author}
  {\bibfnamefont {K.}~\bibnamefont {Rossnagel}}, \bibinfo {author}
  {\bibfnamefont {S.}~\bibnamefont {Sch{\"a}fer}},\ and\ \bibinfo {author}
  {\bibfnamefont {C.}~\bibnamefont {Ropers}},\ }\bibfield  {title} {\bibinfo
  {title} {{Phase ordering of charge density waves traced by ultrafast
  low-energy electron diffraction}},\ }\href
  {https://doi.org/10.1038/nphys4309} {\bibfield  {journal} {\bibinfo
  {journal} {Nat. Phys.}\ }\textbf {\bibinfo {volume} {14}},\ \bibinfo {pages}
  {184} (\bibinfo {year} {2018})}\BibitemShut {NoStop}%
\bibitem [{\citenamefont {Zong}\ \emph {et~al.}(2019)\citenamefont {Zong},
  \citenamefont {Kogar}, \citenamefont {Bie}, \citenamefont {Rohwer},
  \citenamefont {Lee}, \citenamefont {Baldini}, \citenamefont {Erge{\c{c}}en},
  \citenamefont {Yilmaz}, \citenamefont {Freelon}, \citenamefont {Sie},
  \citenamefont {Zhou}, \citenamefont {Straquadine}, \citenamefont {Walmsley},
  \citenamefont {Dolgirev}, \citenamefont {Rozhkov}, \citenamefont {Fisher},
  \citenamefont {Jarillo-Herrero}, \citenamefont {Fine},\ and\ \citenamefont
  {Gedik}}]{Zong2019}%
  \BibitemOpen
  \bibfield  {author} {\bibinfo {author} {\bibfnamefont {A.}~\bibnamefont
  {Zong}}, \bibinfo {author} {\bibfnamefont {A.}~\bibnamefont {Kogar}},
  \bibinfo {author} {\bibfnamefont {Y.-Q.}\ \bibnamefont {Bie}}, \bibinfo
  {author} {\bibfnamefont {T.}~\bibnamefont {Rohwer}}, \bibinfo {author}
  {\bibfnamefont {C.}~\bibnamefont {Lee}}, \bibinfo {author} {\bibfnamefont
  {E.}~\bibnamefont {Baldini}}, \bibinfo {author} {\bibfnamefont
  {E.}~\bibnamefont {Erge{\c{c}}en}}, \bibinfo {author} {\bibfnamefont {M.~B.}\
  \bibnamefont {Yilmaz}}, \bibinfo {author} {\bibfnamefont {B.}~\bibnamefont
  {Freelon}}, \bibinfo {author} {\bibfnamefont {E.~J.}\ \bibnamefont {Sie}},
  \bibinfo {author} {\bibfnamefont {H.}~\bibnamefont {Zhou}}, \bibinfo {author}
  {\bibfnamefont {J.}~\bibnamefont {Straquadine}}, \bibinfo {author}
  {\bibfnamefont {P.}~\bibnamefont {Walmsley}}, \bibinfo {author}
  {\bibfnamefont {P.~E.}\ \bibnamefont {Dolgirev}}, \bibinfo {author}
  {\bibfnamefont {A.~V.}\ \bibnamefont {Rozhkov}}, \bibinfo {author}
  {\bibfnamefont {I.~R.}\ \bibnamefont {Fisher}}, \bibinfo {author}
  {\bibfnamefont {P.}~\bibnamefont {Jarillo-Herrero}}, \bibinfo {author}
  {\bibfnamefont {B.~V.}\ \bibnamefont {Fine}},\ and\ \bibinfo {author}
  {\bibfnamefont {N.}~\bibnamefont {Gedik}},\ }\bibfield  {title} {\bibinfo
  {title} {{Evidence for topological defects in a photoinduced phase
  transition}},\ }\href {https://doi.org/10.1038/s41567-018-0311-9} {\bibfield
  {journal} {\bibinfo  {journal} {Nat. Phys.}\ }\textbf {\bibinfo {volume}
  {15}},\ \bibinfo {pages} {27} (\bibinfo {year} {2019})}\BibitemShut {NoStop}%
\bibitem [{\citenamefont {Laulh{\'{e}}}\ \emph {et~al.}(2017)\citenamefont
  {Laulh{\'{e}}}, \citenamefont {Huber}, \citenamefont {Lantz}, \citenamefont
  {Ferrer}, \citenamefont {Mariager}, \citenamefont {Gr{\"{u}}bel},
  \citenamefont {Rittmann}, \citenamefont {Johnson}, \citenamefont {Esposito},
  \citenamefont {L{\"{u}}bcke}, \citenamefont {Huber}, \citenamefont {Kubli},
  \citenamefont {Savoini}, \citenamefont {Jacques}, \citenamefont {Cario},
  \citenamefont {Corraze}, \citenamefont {Janod}, \citenamefont {Ingold},
  \citenamefont {Beaud}, \citenamefont {Johnson},\ and\ \citenamefont
  {Ravy}}]{Laulhe2017}%
  \BibitemOpen
  \bibfield  {author} {\bibinfo {author} {\bibfnamefont {C.}~\bibnamefont
  {Laulh{\'{e}}}}, \bibinfo {author} {\bibfnamefont {T.}~\bibnamefont {Huber}},
  \bibinfo {author} {\bibfnamefont {G.}~\bibnamefont {Lantz}}, \bibinfo
  {author} {\bibfnamefont {A.}~\bibnamefont {Ferrer}}, \bibinfo {author}
  {\bibfnamefont {S.~O.}\ \bibnamefont {Mariager}}, \bibinfo {author}
  {\bibfnamefont {S.}~\bibnamefont {Gr{\"{u}}bel}}, \bibinfo {author}
  {\bibfnamefont {J.}~\bibnamefont {Rittmann}}, \bibinfo {author}
  {\bibfnamefont {J.~A.}\ \bibnamefont {Johnson}}, \bibinfo {author}
  {\bibfnamefont {V.}~\bibnamefont {Esposito}}, \bibinfo {author}
  {\bibfnamefont {A.}~\bibnamefont {L{\"{u}}bcke}}, \bibinfo {author}
  {\bibfnamefont {L.}~\bibnamefont {Huber}}, \bibinfo {author} {\bibfnamefont
  {M.}~\bibnamefont {Kubli}}, \bibinfo {author} {\bibfnamefont
  {M.}~\bibnamefont {Savoini}}, \bibinfo {author} {\bibfnamefont {V.~L.~R.}\
  \bibnamefont {Jacques}}, \bibinfo {author} {\bibfnamefont {L.}~\bibnamefont
  {Cario}}, \bibinfo {author} {\bibfnamefont {B.}~\bibnamefont {Corraze}},
  \bibinfo {author} {\bibfnamefont {E.}~\bibnamefont {Janod}}, \bibinfo
  {author} {\bibfnamefont {G.}~\bibnamefont {Ingold}}, \bibinfo {author}
  {\bibfnamefont {P.}~\bibnamefont {Beaud}}, \bibinfo {author} {\bibfnamefont
  {S.~L.}\ \bibnamefont {Johnson}},\ and\ \bibinfo {author} {\bibfnamefont
  {S.}~\bibnamefont {Ravy}},\ }\bibfield  {title} {\bibinfo {title} {{Ultrafast
  formation of a charge density wave state in 1$T$-TaS$_2$: Observation at
  nanometer scales using time-resolved X-ray diffraction}},\ }\href
  {https://doi.org/10.1103/PhysRevLett.118.247401} {\bibfield  {journal}
  {\bibinfo  {journal} {Phys. Rev. Lett.}\ }\textbf {\bibinfo {volume} {118}},\
  \bibinfo {pages} {247401} (\bibinfo {year} {2017})}\BibitemShut {NoStop}%
\bibitem [{\citenamefont {Chuang}\ \emph {et~al.}(1991)\citenamefont {Chuang},
  \citenamefont {Durrer}, \citenamefont {Turok},\ and\ \citenamefont
  {Yurke}}]{Chuang1991}%
  \BibitemOpen
  \bibfield  {author} {\bibinfo {author} {\bibfnamefont {I.}~\bibnamefont
  {Chuang}}, \bibinfo {author} {\bibfnamefont {R.}~\bibnamefont {Durrer}},
  \bibinfo {author} {\bibfnamefont {N.}~\bibnamefont {Turok}},\ and\ \bibinfo
  {author} {\bibfnamefont {B.}~\bibnamefont {Yurke}},\ }\bibfield  {title}
  {\bibinfo {title} {{Cosmology in the laboratory: Defect dynamics in liquid
  crystals}},\ }\href {https://doi.org/10.1126/science.251.4999.1336}
  {\bibfield  {journal} {\bibinfo  {journal} {Science}\ }\textbf {\bibinfo
  {volume} {251}},\ \bibinfo {pages} {1336} (\bibinfo {year}
  {1991})}\BibitemShut {NoStop}%
\bibitem [{\citenamefont {Bowick}\ \emph {et~al.}(1994)\citenamefont {Bowick},
  \citenamefont {Chandar}, \citenamefont {Schiff},\ and\ \citenamefont
  {Srivastava}}]{Bowick1994}%
  \BibitemOpen
  \bibfield  {author} {\bibinfo {author} {\bibfnamefont {M.~J.}\ \bibnamefont
  {Bowick}}, \bibinfo {author} {\bibfnamefont {L.}~\bibnamefont {Chandar}},
  \bibinfo {author} {\bibfnamefont {E.~A.}\ \bibnamefont {Schiff}},\ and\
  \bibinfo {author} {\bibfnamefont {A.~M.}\ \bibnamefont {Srivastava}},\
  }\bibfield  {title} {\bibinfo {title} {{The cosmological Kibble mechanism in
  the laboratory: string formation in liquid crystals}},\ }\href
  {https://doi.org/10.1126/science.263.5149.943} {\bibfield  {journal}
  {\bibinfo  {journal} {Science}\ }\textbf {\bibinfo {volume} {263}},\ \bibinfo
  {pages} {943} (\bibinfo {year} {1994})}\BibitemShut {NoStop}%
\bibitem [{\citenamefont {Orihara}\ and\ \citenamefont
  {Ishibashi}(1986)}]{Orihara1986}%
  \BibitemOpen
  \bibfield  {author} {\bibinfo {author} {\bibfnamefont {H.}~\bibnamefont
  {Orihara}}\ and\ \bibinfo {author} {\bibfnamefont {Y.}~\bibnamefont
  {Ishibashi}},\ }\bibfield  {title} {\bibinfo {title} {{Dynamics of
  Disclinations in Twisted Nematics Quenched below the Clearing Point}},\
  }\href {https://doi.org/10.1143/JPSJ.55.2151} {\bibfield  {journal} {\bibinfo
   {journal} {J. Phys. Soc. Jpn.}\ }\textbf {\bibinfo {volume} {55}},\ \bibinfo
  {pages} {2151} (\bibinfo {year} {1986})}\BibitemShut {NoStop}%
\bibitem [{\citenamefont {Nagaya}\ \emph {et~al.}(1992)\citenamefont {Nagaya},
  \citenamefont {Hotta},\ and\ \citenamefont {{Oriharaand Yoshihiro
  Ishibashi}}}]{Nagaya1992}%
  \BibitemOpen
  \bibfield  {author} {\bibinfo {author} {\bibfnamefont {T.}~\bibnamefont
  {Nagaya}}, \bibinfo {author} {\bibfnamefont {H.}~\bibnamefont {Hotta}},\ and\
  \bibinfo {author} {\bibfnamefont {H.}~\bibnamefont {{Oriharaand Yoshihiro
  Ishibashi}}},\ }\bibfield  {title} {\bibinfo {title} {{Experimental study of
  the coarsening dynamics of $+1$ and $-1$ disclinations}},\ }\href
  {https://doi.org/10.1143/JPSJ.61.3511} {\bibfield  {journal} {\bibinfo
  {journal} {J. Phys. Soc. Jpn.}\ }\textbf {\bibinfo {volume} {61}},\ \bibinfo
  {pages} {3511} (\bibinfo {year} {1992})}\BibitemShut {NoStop}%
\bibitem [{\citenamefont {Nakai}\ \emph {et~al.}(1996)\citenamefont {Nakai},
  \citenamefont {Shiwaku}, \citenamefont {Wang}, \citenamefont {Hasegawa},\
  and\ \citenamefont {Hashimoto}}]{Nakai1996}%
  \BibitemOpen
  \bibfield  {author} {\bibinfo {author} {\bibfnamefont {A.}~\bibnamefont
  {Nakai}}, \bibinfo {author} {\bibfnamefont {T.}~\bibnamefont {Shiwaku}},
  \bibinfo {author} {\bibfnamefont {W.}~\bibnamefont {Wang}}, \bibinfo {author}
  {\bibfnamefont {H.}~\bibnamefont {Hasegawa}},\ and\ \bibinfo {author}
  {\bibfnamefont {T.}~\bibnamefont {Hashimoto}},\ }\bibfield  {title} {\bibinfo
  {title} {{Phase-separated structures formed in polymer mixtures containing a
  thermotropic liquid crystalline copolyester as one component}},\ }\href
  {https://doi.org/10.1016/0032-3861(96)85872-1} {\bibfield  {journal}
  {\bibinfo  {journal} {Polymer}\ }\textbf {\bibinfo {volume} {37}},\ \bibinfo
  {pages} {2259} (\bibinfo {year} {1996})}\BibitemShut {NoStop}%
\bibitem [{\citenamefont {Nelson}\ and\ \citenamefont
  {Halperin}(1979)}]{Nelson1979}%
  \BibitemOpen
  \bibfield  {author} {\bibinfo {author} {\bibfnamefont {D.~R.}\ \bibnamefont
  {Nelson}}\ and\ \bibinfo {author} {\bibfnamefont {B.~I.}\ \bibnamefont
  {Halperin}},\ }\bibfield  {title} {\bibinfo {title} {{Dislocation-mediated
  melting in two dimensions}},\ }\href
  {https://doi.org/10.1103/PhysRevB.19.2457} {\bibfield  {journal} {\bibinfo
  {journal} {Phys. Rev. B}\ }\textbf {\bibinfo {volume} {19}},\ \bibinfo
  {pages} {2457} (\bibinfo {year} {1979})}\BibitemShut {NoStop}%
\bibitem [{\citenamefont {Dai}\ \emph {et~al.}(1991)\citenamefont {Dai},
  \citenamefont {Chen},\ and\ \citenamefont {Lieber}}]{Dai1991}%
  \BibitemOpen
  \bibfield  {author} {\bibinfo {author} {\bibfnamefont {H.}~\bibnamefont
  {Dai}}, \bibinfo {author} {\bibfnamefont {H.}~\bibnamefont {Chen}},\ and\
  \bibinfo {author} {\bibfnamefont {C.~M.}\ \bibnamefont {Lieber}},\ }\bibfield
   {title} {\bibinfo {title} {{Weak pinning and hexatic order in a doped
  two-dimensional charge-density-wave system}},\ }\href
  {https://doi.org/10.1103/PhysRevLett.66.3183} {\bibfield  {journal} {\bibinfo
   {journal} {Phys. Rev. Lett.}\ }\textbf {\bibinfo {volume} {66}},\ \bibinfo
  {pages} {3183} (\bibinfo {year} {1991})}\BibitemShut {NoStop}%
\bibitem [{\citenamefont {Soumyanarayanan}\ \emph {et~al.}(2013)\citenamefont
  {Soumyanarayanan}, \citenamefont {Yee}, \citenamefont {He}, \citenamefont
  {van Wezel}, \citenamefont {Rahn}, \citenamefont {Rossnagel}, \citenamefont
  {Hudson}, \citenamefont {Norman},\ and\ \citenamefont
  {Hoffman}}]{Soumyanarayanan2013}%
  \BibitemOpen
  \bibfield  {author} {\bibinfo {author} {\bibfnamefont {A.}~\bibnamefont
  {Soumyanarayanan}}, \bibinfo {author} {\bibfnamefont {M.~M.}\ \bibnamefont
  {Yee}}, \bibinfo {author} {\bibfnamefont {Y.}~\bibnamefont {He}}, \bibinfo
  {author} {\bibfnamefont {J.}~\bibnamefont {van Wezel}}, \bibinfo {author}
  {\bibfnamefont {D.~J.}\ \bibnamefont {Rahn}}, \bibinfo {author}
  {\bibfnamefont {K.}~\bibnamefont {Rossnagel}}, \bibinfo {author}
  {\bibfnamefont {E.~W.}\ \bibnamefont {Hudson}}, \bibinfo {author}
  {\bibfnamefont {M.~R.}\ \bibnamefont {Norman}},\ and\ \bibinfo {author}
  {\bibfnamefont {J.~E.}\ \bibnamefont {Hoffman}},\ }\bibfield  {title}
  {\bibinfo {title} {{Quantum phase transition from triangular to stripe charge
  order in NbSe$_2$}},\ }\href {https://doi.org/10.1073/pnas.1211387110}
  {\bibfield  {journal} {\bibinfo  {journal} {Proc. Natl. Acad. Sci. U.S.A.}\
  }\textbf {\bibinfo {volume} {110}},\ \bibinfo {pages} {1623} (\bibinfo {year}
  {2013})}\BibitemShut {NoStop}%
\bibitem [{\citenamefont {Ma}\ \emph {et~al.}(2016)\citenamefont {Ma},
  \citenamefont {Ye}, \citenamefont {Yu}, \citenamefont {Lu}, \citenamefont
  {Niu}, \citenamefont {Kim}, \citenamefont {Feng}, \citenamefont
  {Tom{\'{a}}nek}, \citenamefont {Son}, \citenamefont {Chen},\ and\
  \citenamefont {Zhang}}]{Ma2016}%
  \BibitemOpen
  \bibfield  {author} {\bibinfo {author} {\bibfnamefont {L.}~\bibnamefont
  {Ma}}, \bibinfo {author} {\bibfnamefont {C.}~\bibnamefont {Ye}}, \bibinfo
  {author} {\bibfnamefont {Y.}~\bibnamefont {Yu}}, \bibinfo {author}
  {\bibfnamefont {X.~F.}\ \bibnamefont {Lu}}, \bibinfo {author} {\bibfnamefont
  {X.}~\bibnamefont {Niu}}, \bibinfo {author} {\bibfnamefont {S.}~\bibnamefont
  {Kim}}, \bibinfo {author} {\bibfnamefont {D.}~\bibnamefont {Feng}}, \bibinfo
  {author} {\bibfnamefont {D.}~\bibnamefont {Tom{\'{a}}nek}}, \bibinfo {author}
  {\bibfnamefont {Y.-W.}\ \bibnamefont {Son}}, \bibinfo {author} {\bibfnamefont
  {X.~H.}\ \bibnamefont {Chen}},\ and\ \bibinfo {author} {\bibfnamefont
  {Y.}~\bibnamefont {Zhang}},\ }\bibfield  {title} {\bibinfo {title} {{A
  metallic mosaic phase and the origin of Mott-insulating state in
  1$T$-TaS$_2$}},\ }\href {https://doi.org/10.1038/ncomms10956} {\bibfield
  {journal} {\bibinfo  {journal} {Nat. Commun.}\ }\textbf {\bibinfo {volume}
  {7}},\ \bibinfo {pages} {10956} (\bibinfo {year} {2016})}\BibitemShut
  {NoStop}%
\bibitem [{\citenamefont {Cho}\ \emph {et~al.}(2016)\citenamefont {Cho},
  \citenamefont {Cheon}, \citenamefont {Kim}, \citenamefont {Lee},
  \citenamefont {Cho}, \citenamefont {Cheong},\ and\ \citenamefont
  {Yeom}}]{Cho2016}%
  \BibitemOpen
  \bibfield  {author} {\bibinfo {author} {\bibfnamefont {D.}~\bibnamefont
  {Cho}}, \bibinfo {author} {\bibfnamefont {S.}~\bibnamefont {Cheon}}, \bibinfo
  {author} {\bibfnamefont {K.-S.}\ \bibnamefont {Kim}}, \bibinfo {author}
  {\bibfnamefont {S.-H.}\ \bibnamefont {Lee}}, \bibinfo {author} {\bibfnamefont
  {Y.-H.}\ \bibnamefont {Cho}}, \bibinfo {author} {\bibfnamefont {S.-W.}\
  \bibnamefont {Cheong}},\ and\ \bibinfo {author} {\bibfnamefont {H.~W.}\
  \bibnamefont {Yeom}},\ }\bibfield  {title} {\bibinfo {title} {{Nanoscale
  manipulation of the Mott insulating state coupled to charge order in
  1$T$-TaS$_2$}},\ }\href {https://doi.org/10.1038/ncomms10453} {\bibfield
  {journal} {\bibinfo  {journal} {Nat. Commun.}\ }\textbf {\bibinfo {volume}
  {7}},\ \bibinfo {pages} {10453} (\bibinfo {year} {2016})}\BibitemShut
  {NoStop}%
\bibitem [{\citenamefont {Chaikin}\ and\ \citenamefont
  {Lubensky}(1995)}]{Chaikin1995}%
  \BibitemOpen
  \bibfield  {author} {\bibinfo {author} {\bibfnamefont {P.~M.}\ \bibnamefont
  {Chaikin}}\ and\ \bibinfo {author} {\bibfnamefont {T.~C.}\ \bibnamefont
  {Lubensky}},\ }\href {https://doi.org/10.1017/CBO9780511813467} {\emph
  {\bibinfo {title} {{Principles of Condensed Matter Physics}}}}\ (\bibinfo
  {publisher} {Cambridge University Press},\ \bibinfo {address} {Cambridge},\
  \bibinfo {year} {1995})\BibitemShut {NoStop}%
\bibitem [{\citenamefont {Qi}\ \emph {et~al.}(2020)\citenamefont {Qi},
  \citenamefont {Ma}, \citenamefont {Zhao}, \citenamefont {Cheng},
  \citenamefont {Jiang}, \citenamefont {Lu}, \citenamefont {Jiang},
  \citenamefont {Qian}, \citenamefont {Wang}, \citenamefont {Zhang},
  \citenamefont {Zhu}, \citenamefont {Zou}, \citenamefont {Wan}, \citenamefont
  {Xiang},\ and\ \citenamefont {Zhang}}]{Qi2020}%
  \BibitemOpen
  \bibfield  {author} {\bibinfo {author} {\bibfnamefont {F.}~\bibnamefont
  {Qi}}, \bibinfo {author} {\bibfnamefont {Z.}~\bibnamefont {Ma}}, \bibinfo
  {author} {\bibfnamefont {L.}~\bibnamefont {Zhao}}, \bibinfo {author}
  {\bibfnamefont {Y.}~\bibnamefont {Cheng}}, \bibinfo {author} {\bibfnamefont
  {W.}~\bibnamefont {Jiang}}, \bibinfo {author} {\bibfnamefont
  {C.}~\bibnamefont {Lu}}, \bibinfo {author} {\bibfnamefont {T.}~\bibnamefont
  {Jiang}}, \bibinfo {author} {\bibfnamefont {D.}~\bibnamefont {Qian}},
  \bibinfo {author} {\bibfnamefont {Z.}~\bibnamefont {Wang}}, \bibinfo {author}
  {\bibfnamefont {W.}~\bibnamefont {Zhang}}, \bibinfo {author} {\bibfnamefont
  {P.}~\bibnamefont {Zhu}}, \bibinfo {author} {\bibfnamefont {X.}~\bibnamefont
  {Zou}}, \bibinfo {author} {\bibfnamefont {W.}~\bibnamefont {Wan}}, \bibinfo
  {author} {\bibfnamefont {D.}~\bibnamefont {Xiang}},\ and\ \bibinfo {author}
  {\bibfnamefont {J.}~\bibnamefont {Zhang}},\ }\bibfield  {title} {\bibinfo
  {title} {{Breaking 50 femtosecond resolution barrier in MeV ultrafast
  electron diffraction with a double bend achromat compressor}},\ }\href
  {https://doi.org/10.1103/PhysRevLett.124.134803} {\bibfield  {journal}
  {\bibinfo  {journal} {Phys. Rev. Lett.}\ }\textbf {\bibinfo {volume} {124}},\
  \bibinfo {pages} {134803} (\bibinfo {year} {2020})}\BibitemShut {NoStop}%
\bibitem [{\citenamefont {Stern}\ \emph {et~al.}(2018)\citenamefont {Stern},
  \citenamefont {{Ren{\'{e}} de Cotret}}, \citenamefont {Otto}, \citenamefont
  {Chatelain}, \citenamefont {Boisvert}, \citenamefont {Sutton},\ and\
  \citenamefont {Siwick}}]{Stern2018}%
  \BibitemOpen
  \bibfield  {author} {\bibinfo {author} {\bibfnamefont {M.~J.}\ \bibnamefont
  {Stern}}, \bibinfo {author} {\bibfnamefont {L.~P.}\ \bibnamefont {{Ren{\'{e}}
  de Cotret}}}, \bibinfo {author} {\bibfnamefont {M.~R.}\ \bibnamefont {Otto}},
  \bibinfo {author} {\bibfnamefont {R.~P.}\ \bibnamefont {Chatelain}}, \bibinfo
  {author} {\bibfnamefont {J.-P.}\ \bibnamefont {Boisvert}}, \bibinfo {author}
  {\bibfnamefont {M.}~\bibnamefont {Sutton}},\ and\ \bibinfo {author}
  {\bibfnamefont {B.~J.}\ \bibnamefont {Siwick}},\ }\bibfield  {title}
  {\bibinfo {title} {{Mapping momentum-dependent electron-phonon coupling and
  nonequilibrium phonon dynamics with ultrafast electron diffuse scattering}},\
  }\href {https://doi.org/10.1103/PhysRevB.97.165416} {\bibfield  {journal}
  {\bibinfo  {journal} {Phys. Rev. B}\ }\textbf {\bibinfo {volume} {97}},\
  \bibinfo {pages} {165416} (\bibinfo {year} {2018})}\BibitemShut {NoStop}%
\bibitem [{\citenamefont {Chen}\ \emph {et~al.}(2016)\citenamefont {Chen},
  \citenamefont {Chan}, \citenamefont {Fang}, \citenamefont {Mo}, \citenamefont
  {Hussain}, \citenamefont {Fedorov}, \citenamefont {Chou},\ and\ \citenamefont
  {Chiang}}]{Chen2016}%
  \BibitemOpen
  \bibfield  {author} {\bibinfo {author} {\bibfnamefont {P.}~\bibnamefont
  {Chen}}, \bibinfo {author} {\bibfnamefont {Y.-H.}\ \bibnamefont {Chan}},
  \bibinfo {author} {\bibfnamefont {X.-Y.}\ \bibnamefont {Fang}}, \bibinfo
  {author} {\bibfnamefont {S.-K.}\ \bibnamefont {Mo}}, \bibinfo {author}
  {\bibfnamefont {Z.}~\bibnamefont {Hussain}}, \bibinfo {author} {\bibfnamefont
  {A.-V.}\ \bibnamefont {Fedorov}}, \bibinfo {author} {\bibfnamefont {M.~Y.}\
  \bibnamefont {Chou}},\ and\ \bibinfo {author} {\bibfnamefont {T.-C.}\
  \bibnamefont {Chiang}},\ }\bibfield  {title} {\bibinfo {title} {{Hidden order
  and dimensional crossover of the charge density waves in TiSe$_2$}},\ }\href
  {https://doi.org/10.1038/srep37910} {\bibfield  {journal} {\bibinfo
  {journal} {Sci. Rep.}\ }\textbf {\bibinfo {volume} {6}},\ \bibinfo {pages}
  {37910} (\bibinfo {year} {2016})}\BibitemShut {NoStop}%
\bibitem [{\citenamefont {Cheng}\ \emph {et~al.}(2022)\citenamefont {Cheng},
  \citenamefont {Zong}, \citenamefont {Li}, \citenamefont {Xia}, \citenamefont
  {Duan}, \citenamefont {Zhao}, \citenamefont {Li}, \citenamefont {Qi},
  \citenamefont {Wu}, \citenamefont {Zhao}, \citenamefont {Zhu}, \citenamefont
  {Zou}, \citenamefont {Jiang}, \citenamefont {Guo}, \citenamefont {Yang},
  \citenamefont {Qian}, \citenamefont {Zhang}, \citenamefont {Kogar},
  \citenamefont {Zuerch}, \citenamefont {Xiang},\ and\ \citenamefont
  {Zhang}}]{Cheng2022}%
  \BibitemOpen
  \bibfield  {author} {\bibinfo {author} {\bibfnamefont {Y.}~\bibnamefont
  {Cheng}}, \bibinfo {author} {\bibfnamefont {A.}~\bibnamefont {Zong}},
  \bibinfo {author} {\bibfnamefont {J.}~\bibnamefont {Li}}, \bibinfo {author}
  {\bibfnamefont {W.}~\bibnamefont {Xia}}, \bibinfo {author} {\bibfnamefont
  {S.}~\bibnamefont {Duan}}, \bibinfo {author} {\bibfnamefont {W.}~\bibnamefont
  {Zhao}}, \bibinfo {author} {\bibfnamefont {Y.}~\bibnamefont {Li}}, \bibinfo
  {author} {\bibfnamefont {F.}~\bibnamefont {Qi}}, \bibinfo {author}
  {\bibfnamefont {J.}~\bibnamefont {Wu}}, \bibinfo {author} {\bibfnamefont
  {L.}~\bibnamefont {Zhao}}, \bibinfo {author} {\bibfnamefont {P.}~\bibnamefont
  {Zhu}}, \bibinfo {author} {\bibfnamefont {X.}~\bibnamefont {Zou}}, \bibinfo
  {author} {\bibfnamefont {T.}~\bibnamefont {Jiang}}, \bibinfo {author}
  {\bibfnamefont {Y.}~\bibnamefont {Guo}}, \bibinfo {author} {\bibfnamefont
  {L.}~\bibnamefont {Yang}}, \bibinfo {author} {\bibfnamefont {D.}~\bibnamefont
  {Qian}}, \bibinfo {author} {\bibfnamefont {W.}~\bibnamefont {Zhang}},
  \bibinfo {author} {\bibfnamefont {A.}~\bibnamefont {Kogar}}, \bibinfo
  {author} {\bibfnamefont {M.~W.}\ \bibnamefont {Zuerch}}, \bibinfo {author}
  {\bibfnamefont {D.}~\bibnamefont {Xiang}},\ and\ \bibinfo {author}
  {\bibfnamefont {J.}~\bibnamefont {Zhang}},\ }\bibfield  {title} {\bibinfo
  {title} {{Light-induced dimension crossover dictated by excitonic
  correlations}},\ }\href {https://doi.org/10.1038/s41467-022-28309-5}
  {\bibfield  {journal} {\bibinfo  {journal} {Nat. Commun.}\ }\textbf {\bibinfo
  {volume} {13}},\ \bibinfo {pages} {963} (\bibinfo {year} {2022})}\BibitemShut
  {NoStop}%
\bibitem [{SM()}]{SM}%
  \BibitemOpen
  \href@noop {} {}\bibinfo {note} {See supplemental materials.}\BibitemShut
  {Stop}%
\bibitem [{Rem()}]{Remark1}%
  \BibitemOpen
  \href@noop {} {}\bibinfo {note} {In principle, peak broadening may result
  from either amplitude or phase fluctuations in the CDW, the latter of which
  are associated with topological defects due to the commensurate nature of the
  CDW. In 1$T$-TiSe$_2$, the relation 2$\Delta/k_B T_c\gg 3.53$ holds down to
  the monolayer limit \cite{Watson2020}, where $k_B$ is the Boltzmann constant
  and $\Delta$ is the CDW energy gap in the electronic dispersion. Hence, the
  CDW in 1$T$-TiSe$_2$ is in the strong-coupling regime, and we expect phase
  fluctuations to dominate, justifying the relation between topological defects
  and the CDW peak width.}\BibitemShut {Stop}%
\bibitem [{\citenamefont {Zhu}\ \emph {et~al.}(2015)\citenamefont {Zhu},
  \citenamefont {Zhu}, \citenamefont {Hidaka}, \citenamefont {Wu},
  \citenamefont {Cao}, \citenamefont {Berger}, \citenamefont {Geck},
  \citenamefont {Kraus}, \citenamefont {Pjerov}, \citenamefont {Shen},
  \citenamefont {Tobey}, \citenamefont {Hill},\ and\ \citenamefont
  {Wang}}]{Zhu2015}%
  \BibitemOpen
  \bibfield  {author} {\bibinfo {author} {\bibfnamefont {P.}~\bibnamefont
  {Zhu}}, \bibinfo {author} {\bibfnamefont {Y.}~\bibnamefont {Zhu}}, \bibinfo
  {author} {\bibfnamefont {Y.}~\bibnamefont {Hidaka}}, \bibinfo {author}
  {\bibfnamefont {L.}~\bibnamefont {Wu}}, \bibinfo {author} {\bibfnamefont
  {J.}~\bibnamefont {Cao}}, \bibinfo {author} {\bibfnamefont {H.}~\bibnamefont
  {Berger}}, \bibinfo {author} {\bibfnamefont {J.}~\bibnamefont {Geck}},
  \bibinfo {author} {\bibfnamefont {R.}~\bibnamefont {Kraus}}, \bibinfo
  {author} {\bibfnamefont {S.}~\bibnamefont {Pjerov}}, \bibinfo {author}
  {\bibfnamefont {Y.}~\bibnamefont {Shen}}, \bibinfo {author} {\bibfnamefont
  {R.~I.}\ \bibnamefont {Tobey}}, \bibinfo {author} {\bibfnamefont {J.~P.}\
  \bibnamefont {Hill}},\ and\ \bibinfo {author} {\bibfnamefont {X.~J.}\
  \bibnamefont {Wang}},\ }\bibfield  {title} {\bibinfo {title} {{Femtosecond
  time-resolved MeV electron diffraction}},\ }\href
  {https://doi.org/10.1088/1367-2630/17/6/063004} {\bibfield  {journal}
  {\bibinfo  {journal} {New J. Phys.}\ }\textbf {\bibinfo {volume} {17}},\
  \bibinfo {pages} {063004} (\bibinfo {year} {2015})}\BibitemShut {NoStop}%
\bibitem [{\citenamefont {Otto}\ \emph {et~al.}(2021)\citenamefont {Otto},
  \citenamefont {P{\"{o}}hls}, \citenamefont {{Ren{\'{e}} de Cotret}},
  \citenamefont {Stern}, \citenamefont {Sutton},\ and\ \citenamefont
  {Siwick}}]{Otto2021}%
  \BibitemOpen
  \bibfield  {author} {\bibinfo {author} {\bibfnamefont {M.~R.}\ \bibnamefont
  {Otto}}, \bibinfo {author} {\bibfnamefont {J.-H.}\ \bibnamefont
  {P{\"{o}}hls}}, \bibinfo {author} {\bibfnamefont {L.~P.}\ \bibnamefont
  {{Ren{\'{e}} de Cotret}}}, \bibinfo {author} {\bibfnamefont {M.~J.}\
  \bibnamefont {Stern}}, \bibinfo {author} {\bibfnamefont {M.}~\bibnamefont
  {Sutton}},\ and\ \bibinfo {author} {\bibfnamefont {B.~J.}\ \bibnamefont
  {Siwick}},\ }\bibfield  {title} {\bibinfo {title} {{Mechanisms of
  electron-phonon coupling unraveled in momentum and time: The case of soft
  phonons in TiSe$_2$}},\ }\href {https://doi.org/10.1126/sciadv.abf2810}
  {\bibfield  {journal} {\bibinfo  {journal} {Sci. Adv.}\ }\textbf {\bibinfo
  {volume} {7}},\ \bibinfo {pages} {eabf2810} (\bibinfo {year}
  {2021})}\BibitemShut {NoStop}%
\bibitem [{\citenamefont {Domr{\"{o}}se}\ \emph {et~al.}(2022)\citenamefont
  {Domr{\"{o}}se}, \citenamefont {Danz}, \citenamefont {Schaible},
  \citenamefont {Rossnagel}, \citenamefont {Yalunin},\ and\ \citenamefont
  {Ropers}}]{Domrose2022}%
  \BibitemOpen
  \bibfield  {author} {\bibinfo {author} {\bibfnamefont {T.}~\bibnamefont
  {Domr{\"{o}}se}}, \bibinfo {author} {\bibfnamefont {T.}~\bibnamefont {Danz}},
  \bibinfo {author} {\bibfnamefont {S.~F.}\ \bibnamefont {Schaible}}, \bibinfo
  {author} {\bibfnamefont {K.}~\bibnamefont {Rossnagel}}, \bibinfo {author}
  {\bibfnamefont {S.~V.}\ \bibnamefont {Yalunin}},\ and\ \bibinfo {author}
  {\bibfnamefont {C.}~\bibnamefont {Ropers}},\ }\href@noop {} {\bibinfo {title}
  {{Light-induced hexatic state in a layered quantum material}}} (\bibinfo
  {year} {2022}),\ \Eprint {https://arxiv.org/abs/2207.05571}
  {arXiv:2207.05571} \BibitemShut {NoStop}%
\bibitem [{\citenamefont {Tarkhov}\ \emph {et~al.}(2022)\citenamefont
  {Tarkhov}, \citenamefont {Rozhkov},\ and\ \citenamefont
  {Fine}}]{Tarkhov2022}%
  \BibitemOpen
  \bibfield  {author} {\bibinfo {author} {\bibfnamefont {A.~E.}\ \bibnamefont
  {Tarkhov}}, \bibinfo {author} {\bibfnamefont {A.~V.}\ \bibnamefont
  {Rozhkov}},\ and\ \bibinfo {author} {\bibfnamefont {B.~V.}\ \bibnamefont
  {Fine}},\ }\bibfield  {title} {\bibinfo {title} {{Dynamics of topological
  defects after photoinduced melting of a charge density wave}},\ }\href
  {https://doi.org/10.1103/PhysRevB.106.L121109} {\bibfield  {journal}
  {\bibinfo  {journal} {Phys. Rev. B}\ }\textbf {\bibinfo {volume} {106}},\
  \bibinfo {pages} {L121109} (\bibinfo {year} {2022})}\BibitemShut {NoStop}%
\bibitem [{\citenamefont {van Wezel}\ \emph {et~al.}(2010)\citenamefont {van
  Wezel}, \citenamefont {Nahai-Williamson},\ and\ \citenamefont
  {Saxena}}]{VanWezel2010}%
  \BibitemOpen
  \bibfield  {author} {\bibinfo {author} {\bibfnamefont {J.}~\bibnamefont {van
  Wezel}}, \bibinfo {author} {\bibfnamefont {P.}~\bibnamefont
  {Nahai-Williamson}},\ and\ \bibinfo {author} {\bibfnamefont {S.~S.}\
  \bibnamefont {Saxena}},\ }\bibfield  {title} {\bibinfo {title}
  {{Exciton-phonon-driven charge density wave in TiSe$_2$}},\ }\href
  {https://doi.org/10.1103/PhysRevB.81.165109} {\bibfield  {journal} {\bibinfo
  {journal} {Phys. Rev. B}\ }\textbf {\bibinfo {volume} {81}},\ \bibinfo
  {pages} {165109} (\bibinfo {year} {2010})}\BibitemShut {NoStop}%
\bibitem [{\citenamefont {Porer}\ \emph {et~al.}(2014)\citenamefont {Porer},
  \citenamefont {Leierseder}, \citenamefont {M{\'{e}}nard}, \citenamefont
  {Dachraoui}, \citenamefont {Mouchliadis}, \citenamefont {Perakis},
  \citenamefont {Heinzmann}, \citenamefont {Demsar}, \citenamefont
  {Rossnagel},\ and\ \citenamefont {Huber}}]{Porer2014}%
  \BibitemOpen
  \bibfield  {author} {\bibinfo {author} {\bibfnamefont {M.}~\bibnamefont
  {Porer}}, \bibinfo {author} {\bibfnamefont {U.}~\bibnamefont {Leierseder}},
  \bibinfo {author} {\bibfnamefont {J.-M.}\ \bibnamefont {M{\'{e}}nard}},
  \bibinfo {author} {\bibfnamefont {H.}~\bibnamefont {Dachraoui}}, \bibinfo
  {author} {\bibfnamefont {L.}~\bibnamefont {Mouchliadis}}, \bibinfo {author}
  {\bibfnamefont {I.~E.}\ \bibnamefont {Perakis}}, \bibinfo {author}
  {\bibfnamefont {U.}~\bibnamefont {Heinzmann}}, \bibinfo {author}
  {\bibfnamefont {J.}~\bibnamefont {Demsar}}, \bibinfo {author} {\bibfnamefont
  {K.}~\bibnamefont {Rossnagel}},\ and\ \bibinfo {author} {\bibfnamefont
  {R.}~\bibnamefont {Huber}},\ }\bibfield  {title} {\bibinfo {title}
  {{Non-thermal separation of electronic and structural orders in a persisting
  charge density wave}},\ }\href {https://doi.org/10.1038/nmat4042} {\bibfield
  {journal} {\bibinfo  {journal} {Nat. Mater.}\ }\textbf {\bibinfo {volume}
  {13}},\ \bibinfo {pages} {857} (\bibinfo {year} {2014})}\BibitemShut
  {NoStop}%
\bibitem [{\citenamefont {Zong}\ \emph {et~al.}(2021)\citenamefont {Zong},
  \citenamefont {Dolgirev}, \citenamefont {Kogar}, \citenamefont {Su},
  \citenamefont {Shen}, \citenamefont {Straquadine}, \citenamefont {Wang},
  \citenamefont {Luo}, \citenamefont {Kozina}, \citenamefont {Reid},
  \citenamefont {Li}, \citenamefont {Yang}, \citenamefont {Weathersby},
  \citenamefont {Park}, \citenamefont {Sie}, \citenamefont {Jarillo-Herrero},
  \citenamefont {Fisher}, \citenamefont {Wang}, \citenamefont {Demler},\ and\
  \citenamefont {Gedik}}]{Zong2021}%
  \BibitemOpen
  \bibfield  {author} {\bibinfo {author} {\bibfnamefont {A.}~\bibnamefont
  {Zong}}, \bibinfo {author} {\bibfnamefont {P.~E.}\ \bibnamefont {Dolgirev}},
  \bibinfo {author} {\bibfnamefont {A.}~\bibnamefont {Kogar}}, \bibinfo
  {author} {\bibfnamefont {Y.}~\bibnamefont {Su}}, \bibinfo {author}
  {\bibfnamefont {X.}~\bibnamefont {Shen}}, \bibinfo {author} {\bibfnamefont
  {J.~A.~W.}\ \bibnamefont {Straquadine}}, \bibinfo {author} {\bibfnamefont
  {X.}~\bibnamefont {Wang}}, \bibinfo {author} {\bibfnamefont {D.}~\bibnamefont
  {Luo}}, \bibinfo {author} {\bibfnamefont {M.~E.}\ \bibnamefont {Kozina}},
  \bibinfo {author} {\bibfnamefont {A.~H.}\ \bibnamefont {Reid}}, \bibinfo
  {author} {\bibfnamefont {R.}~\bibnamefont {Li}}, \bibinfo {author}
  {\bibfnamefont {J.}~\bibnamefont {Yang}}, \bibinfo {author} {\bibfnamefont
  {S.~P.}\ \bibnamefont {Weathersby}}, \bibinfo {author} {\bibfnamefont
  {S.}~\bibnamefont {Park}}, \bibinfo {author} {\bibfnamefont {E.~J.}\
  \bibnamefont {Sie}}, \bibinfo {author} {\bibfnamefont {P.}~\bibnamefont
  {Jarillo-Herrero}}, \bibinfo {author} {\bibfnamefont {I.~R.}\ \bibnamefont
  {Fisher}}, \bibinfo {author} {\bibfnamefont {X.}~\bibnamefont {Wang}},
  \bibinfo {author} {\bibfnamefont {E.}~\bibnamefont {Demler}},\ and\ \bibinfo
  {author} {\bibfnamefont {N.}~\bibnamefont {Gedik}},\ }\bibfield  {title}
  {\bibinfo {title} {{Role of equilibrium fluctuations in light-induced
  order}},\ }\href {https://doi.org/10.1103/PhysRevLett.127.227401} {\bibfield
  {journal} {\bibinfo  {journal} {Phys. Rev. Lett.}\ }\textbf {\bibinfo
  {volume} {127}},\ \bibinfo {pages} {227401} (\bibinfo {year}
  {2021})}\BibitemShut {NoStop}%
\bibitem [{\citenamefont {Wall}\ \emph {et~al.}(2018)\citenamefont {Wall},
  \citenamefont {Yang}, \citenamefont {Vidas}, \citenamefont {Chollet},
  \citenamefont {Glownia}, \citenamefont {Kozina}, \citenamefont {Katayama},
  \citenamefont {Henighan}, \citenamefont {Jiang}, \citenamefont {Miller},
  \citenamefont {Reis}, \citenamefont {Boatner}, \citenamefont {Delaire},\ and\
  \citenamefont {Trigo}}]{Wall2018}%
  \BibitemOpen
  \bibfield  {author} {\bibinfo {author} {\bibfnamefont {S.}~\bibnamefont
  {Wall}}, \bibinfo {author} {\bibfnamefont {S.}~\bibnamefont {Yang}}, \bibinfo
  {author} {\bibfnamefont {L.}~\bibnamefont {Vidas}}, \bibinfo {author}
  {\bibfnamefont {M.}~\bibnamefont {Chollet}}, \bibinfo {author} {\bibfnamefont
  {J.~M.}\ \bibnamefont {Glownia}}, \bibinfo {author} {\bibfnamefont
  {M.}~\bibnamefont {Kozina}}, \bibinfo {author} {\bibfnamefont
  {T.}~\bibnamefont {Katayama}}, \bibinfo {author} {\bibfnamefont
  {T.}~\bibnamefont {Henighan}}, \bibinfo {author} {\bibfnamefont
  {M.}~\bibnamefont {Jiang}}, \bibinfo {author} {\bibfnamefont {T.~A.}\
  \bibnamefont {Miller}}, \bibinfo {author} {\bibfnamefont {D.~A.}\
  \bibnamefont {Reis}}, \bibinfo {author} {\bibfnamefont {L.~A.}\ \bibnamefont
  {Boatner}}, \bibinfo {author} {\bibfnamefont {O.}~\bibnamefont {Delaire}},\
  and\ \bibinfo {author} {\bibfnamefont {M.}~\bibnamefont {Trigo}},\ }\bibfield
   {title} {\bibinfo {title} {{Ultrafast disordering of vanadium dimers in
  photoexcited VO$_2$}},\ }\href {https://doi.org/10.1126/science.aau3873}
  {\bibfield  {journal} {\bibinfo  {journal} {Science}\ }\textbf {\bibinfo
  {volume} {362}},\ \bibinfo {pages} {572} (\bibinfo {year}
  {2018})}\BibitemShut {NoStop}%
\bibitem [{\citenamefont {Xu}\ and\ \citenamefont {Chiang}(2005)}]{Xu2005}%
  \BibitemOpen
  \bibfield  {author} {\bibinfo {author} {\bibfnamefont {R.}~\bibnamefont
  {Xu}}\ and\ \bibinfo {author} {\bibfnamefont {T.~C.}\ \bibnamefont
  {Chiang}},\ }\bibfield  {title} {\bibinfo {title} {{Determination of phonon
  dispersion relations by X-ray thermal diffuse scattering}},\ }\href
  {https://doi.org/10.1524/zkri.2005.220.12.1009} {\bibfield  {journal}
  {\bibinfo  {journal} {Z. Kristallogr. Cryst. Mater.}\ }\textbf {\bibinfo
  {volume} {220}},\ \bibinfo {pages} {1009} (\bibinfo {year}
  {2005})}\BibitemShut {NoStop}%
\bibitem [{\citenamefont {de~la Torre}\ \emph {et~al.}(2021)\citenamefont
  {de~la Torre}, \citenamefont {Kennes}, \citenamefont {Claassen},
  \citenamefont {Gerber}, \citenamefont {McIver},\ and\ \citenamefont
  {Sentef}}]{DelaTorre2021}%
  \BibitemOpen
  \bibfield  {author} {\bibinfo {author} {\bibfnamefont {A.}~\bibnamefont
  {de~la Torre}}, \bibinfo {author} {\bibfnamefont {D.~M.}\ \bibnamefont
  {Kennes}}, \bibinfo {author} {\bibfnamefont {M.}~\bibnamefont {Claassen}},
  \bibinfo {author} {\bibfnamefont {S.}~\bibnamefont {Gerber}}, \bibinfo
  {author} {\bibfnamefont {J.~W.}\ \bibnamefont {McIver}},\ and\ \bibinfo
  {author} {\bibfnamefont {M.~A.}\ \bibnamefont {Sentef}},\ }\bibfield  {title}
  {\bibinfo {title} {{Colloquium: Nonthermal pathways to ultrafast control in
  quantum materials}},\ }\href {https://doi.org/10.1103/RevModPhys.93.041002}
  {\bibfield  {journal} {\bibinfo  {journal} {Rev. Mod. Phys.}\ }\textbf
  {\bibinfo {volume} {93}},\ \bibinfo {pages} {041002} (\bibinfo {year}
  {2021})}\BibitemShut {NoStop}%
\bibitem [{\citenamefont {Watson}\ \emph {et~al.}(2020)\citenamefont {Watson},
  \citenamefont {Rajan}, \citenamefont {Antonelli}, \citenamefont {Underwood},
  \citenamefont {Markovi{\'{c}}}, \citenamefont {Mazzola}, \citenamefont
  {Clark}, \citenamefont {Siemann}, \citenamefont {Biswas}, \citenamefont
  {Hunter}, \citenamefont {Jandura}, \citenamefont {Reichstetter},
  \citenamefont {McLaren}, \citenamefont {{Le F{\`{e}}vre}}, \citenamefont
  {Vinai},\ and\ \citenamefont {King}}]{Watson2020}%
  \BibitemOpen
  \bibfield  {author} {\bibinfo {author} {\bibfnamefont {M.~D.}\ \bibnamefont
  {Watson}}, \bibinfo {author} {\bibfnamefont {A.}~\bibnamefont {Rajan}},
  \bibinfo {author} {\bibfnamefont {T.}~\bibnamefont {Antonelli}}, \bibinfo
  {author} {\bibfnamefont {K.}~\bibnamefont {Underwood}}, \bibinfo {author}
  {\bibfnamefont {I.}~\bibnamefont {Markovi{\'{c}}}}, \bibinfo {author}
  {\bibfnamefont {F.}~\bibnamefont {Mazzola}}, \bibinfo {author} {\bibfnamefont
  {O.~J.}\ \bibnamefont {Clark}}, \bibinfo {author} {\bibfnamefont {G.-R.}\
  \bibnamefont {Siemann}}, \bibinfo {author} {\bibfnamefont {D.}~\bibnamefont
  {Biswas}}, \bibinfo {author} {\bibfnamefont {A.}~\bibnamefont {Hunter}},
  \bibinfo {author} {\bibfnamefont {S.}~\bibnamefont {Jandura}}, \bibinfo
  {author} {\bibfnamefont {J.}~\bibnamefont {Reichstetter}}, \bibinfo {author}
  {\bibfnamefont {M.}~\bibnamefont {McLaren}}, \bibinfo {author} {\bibfnamefont
  {P.}~\bibnamefont {{Le F{\`{e}}vre}}}, \bibinfo {author} {\bibfnamefont
  {G.}~\bibnamefont {Vinai}},\ and\ \bibinfo {author} {\bibfnamefont
  {P.~D.~C.}\ \bibnamefont {King}},\ }\bibfield  {title} {\bibinfo {title}
  {{Strong-coupling charge density wave in monolayer TiSe$_2$}},\ }\href
  {https://doi.org/10.1088/2053-1583/abafec} {\bibfield  {journal} {\bibinfo
  {journal} {2D Mater.}\ }\textbf {\bibinfo {volume} {8}},\ \bibinfo {pages}
  {015004} (\bibinfo {year} {2020})}\BibitemShut {NoStop}%
\end{thebibliography}

%

\end{document}